\documentclass[]{pasj02} 
\usepackage[switch,mathlines]{lineno} 

\jyear{2026}
\Received{2026/10/07}
\Accepted{2026/11/19}

\begin{document} 
\title{An ALMA Survey of Molecular Gas in the Host Galaxies of Fast Radio Bursts}

\author{
Bunyo \textsc{Hatsukade},\altaffilmark{1,2,3}\altemailmark\orcid{0000-0001-6469-8725} \email{bunyo.hatsukade@nao.ac.jp} 
Tetsuya \textsc{Hashimoto},\altaffilmark{4}\orcid{0000-0001-7228-1428}
Yuu \textsc{Niino},\altaffilmark{5}\orcid{0000-0001-5322-5076}
Fumi \textsc{Egusa},\altaffilmark{6}\orcid{0000-0002-1639-1515}
Tzu-Yin \textsc{Hsu},\altaffilmark{7,8}\orcid{0000-0002-0944-5634}
Kotaro \textsc{Kohno},\altaffilmark{6,9}\orcid{0000-0002-4052-2394}
Kouji \textsc{Ohta},\altaffilmark{10}\orcid{0000-0003-3844-1517}
}
\altaffiltext{1}{National Astronomical Observatory of Japan, 2-21-1 Osawa, Mitaka, Tokyo 181-8588, Japan}
\altaffiltext{2}{Graduate University for Advanced Studies, SOKENDAI, Osawa, Mitaka, Tokyo 181-8588, Japan}
\altaffiltext{3}{Department of Astronomy, Graduate School of Science, The University of Tokyo, 7-3-1 Hongo, Bunkyo-ku, Tokyo 133-0033, Japan}
\altaffiltext{4}{Department of Physics, National Chung Hsing University, 145 Xingda Rd., South Dist., Taichung 40227, Taiwan}
\altaffiltext{5}{Kiso Observatory, Institute of Astronomy, Graduate School of Science, The University of Tokyo, 10762-30 Mitake, Kiso-machi, Kiso-gun, Nagano 397-0101, Japan}
\altaffiltext{6}{Institute of Astronomy, The University of Tokyo, 2-21-1 Osawa, Mitaka, Tokyo 181-0015, Japan}
\altaffiltext{7}{Centre for Astrophysics and Supercomputing, Swinburne University of Technology, Hawthorn VIC 3122, Australia}
\altaffiltext{8}{OzGrav: The ARC Centre of Excellence for Gravitational Wave Discovery, Swinburne University of Technology, Hawthorn, VIC 3122, Australia}
\altaffiltext{9}{Research Center for the Early Universe, Graduate School of Science, The University of Tokyo, 7-3-1 Hongo, Bunkyo-ku, Tokyo 113-0033, Japan}
\altaffiltext{10}{Department of Astronomy, Kyoto University, Kyoto, 606-8502, Japan}
\KeyWords{galaxies: ISM --- radio lines: galaxies --- radio lines: ISM}  

\maketitle

\begin{abstract}
Cold gas in fast radio burst (FRB) host galaxies probes the environments where FRB progenitors form.
We present Atacama Large Millimeter/submillimeter Array (ALMA) observations of CO(1--0) and CO(2--1) emission and millimeter continuum measurements for nine localized FRB hosts at $z=0.05$--0.38.
CO emission is significantly detected in six hosts and tentatively detected in one, yielding molecular gas masses of $2.2\times10^8$--$1.0\times10^{10}$ $M_\odot$ for a Milky Way CO-to-H$_2$ conversion factor.
All six detections show spatially resolved emission and large-scale velocity gradients, several consistent with ordered disk rotation.
FRB positions do not necessarily coincide with the brightest CO emission, and no distinct kinematic component is identified at the burst positions at the present kpc-scale resolution.
Together with published data, we assemble the largest FRB-host sample with CO measurements or upper limits to date.
CO-detected hosts tend to have larger molecular gas reservoirs and longer depletion times than stellar-mass- and star-formation-rate-matched xCOLD GASS galaxies.
They also tend to have higher molecular gas fractions and longer depletion times than expected from molecular-gas scaling relations.
However, survival analyses including non-detections do not establish a significant difference in molecular gas mass or gas fraction, so the evidence for longer depletion times is marginal and depends on the test employed.
Both hosts detected in CO and atomic hydrogen have molecular-to-atomic mass ratios within the broad scatter of comparison galaxies.
Our results highlight diverse cold-gas environments rather than a uniformly molecular-gas-rich FRB host population.
At the position of the repeating FRB\,20201124A, the measured 98 GHz continuum flux density is above the predicted star-formation contribution and lies at the lower end of or below an extrapolation of the VLA spectrum.
These results allow a compact persistent-source contribution and a possible spectral turnover between centimeter and millimeter wavelengths, although the present sensitivity and resolution prevent a firm separation from surrounding star-forming emission.
\end{abstract}

\section{Introduction}

Fast radio bursts (FRBs) are luminous radio pulses with durations of microseconds to milliseconds and an extragalactic origin established through their large dispersion measures and host-galaxy associations (e.g., \cite{cord19,hein20a}). 
Since the discovery of the first FRB \citep{lori07}, a wide range of emission and progenitor models have been proposed \citep{plat19,zhan20}. 
The detection of an FRB-like burst from the Galactic magnetar SGR\,1935+2154 demonstrated that magnetars can produce such emission \citep{the20a,boch20}. 
However, it remains unclear how FRB sources are formed, whether a single evolutionary pathway can account for the observed population, and whether repeating and apparently non-repeating sources have different progenitors. 
The broad and overlapping host-galaxy properties of repeaters and apparent non-repeaters have not yet established a clear population division 
\citep{hash20,pleu21,gord23,siro26}.

FRB host galaxies provide constraints on progenitor formation channels and delay times that cannot be obtained from burst properties alone. 
FRB hosts span broad ranges of properties, such as morphology, color, stellar mass, and star formation rate (SFR) \citep{hein20a,bhan22,gord23}.
In the uniformly modeled sample of 23 securely associated hosts presented by \citet{gord23}, 20 (87\%) are classified as star-forming, two as transitioning, and one as quiescent. 
The predominance of star-forming hosts is consistent with a substantial contribution from young magnetars formed through massive-star evolution. 
At the same time, the localization of FRB\,20200120E to an old globular cluster and the discovery of the repeating FRB\,20240209A at a projected offset of approximately 40 kpc from a massive quiescent elliptical galaxy demonstrate that delayed or dynamical formation channels also operate \citep{kirs22,shah25,efte25}. 
Population-level studies similarly yield a more complex picture. 
The deficit of low-mass hosts reported by \citet{shar24} was interpreted as a possible preference for metal-rich environments and binary-mediated magnetar formation, whereas \citet{yama26} found no clear metallicity preference. 
More recent modeling indicates that a mixture of SFR- and stellar-mass-linked components describes the host population better than a population that follows ongoing star formation alone \citep{horo26}. 
These results motivate observations that distinguish the present-day star-forming fuel from the time-integrated stellar population.

Cold gas provides a direct probe of this connection. 
CO emission traces the molecular reservoir available for star formation and constrains molecular gas mass, gas fraction, depletion time, spatial distribution, and kinematics. 
Previous studies found substantial diversity among FRB hosts, including non-detections, systems with molecular-gas properties similar to ordinary star-forming galaxies, and gas-rich hosts with long depletion times \citep{bowe18,hats22,chit23}. 
Spatially resolved CO observations have also revealed contrasting kinematic environments: the FRB\,20180924B host shows an asymmetric velocity gradient \citep{hsu23}, whereas the FRB\,20191001A host contains an ordered rotating molecular disk, despite having a nearby gas-rich companion \citep{yama24a}. 
Atomic hydrogen complements CO by tracing the extended cold-gas reservoir and potential signatures of accretion or interaction 
\citep{mich21,kaur22,lee-23,glow23,glow24,kaur26}.
An expanded \textsc{Hi} host sample found substantial atomic-gas reservoirs but no statistically significant excess of disturbed morphologies relative to normal galaxies \citep{roxb25}. 
The available CO samples remain small and heterogeneous in transition, sensitivity, and spatial resolution. 
It is therefore uncertain whether FRB hosts are systematically molecular-gas rich after controlling for stellar mass and SFR, whether they have unusual star-formation efficiencies, whether their gas kinematics commonly indicate interactions, and whether FRB positions coincide with the brightest molecular-gas regions.

In this paper, we present CO observations of nine localized, spectroscopically confirmed FRB host galaxies at $z=0.05$--0.38 
with the Atacama Large Millimeter/submillimeter Array (ALMA).
We enlarge the sample of FRB hosts with molecular-gas constraints and compare their molecular gas masses, gas fractions, and depletion times with stellar-mass- and SFR-matched galaxies and with molecular-gas scaling relations. 
By including CO non-detections in survival analyses, we test whether any apparent gas enhancement is a population-wide property rather than a selection effect favoring CO-bright hosts. 
We also compare the atomic and molecular gas reservoirs where both are available, and investigate the millimeter continuum emission associated with the repeating FRB\,20201124A. 
Section~\ref{sec:data} describes the sample, observations, and data reduction. 
Section~\ref{sec:results} presents the CO and continuum results. 
Section~\ref{sec:gas} compares the cold-gas properties with those of other galaxies and discusses implications for FRB progenitors, while Section~\ref{sec:prs} examines the radio source associated with FRB\,20201124A. 
Our conclusions are summarized in Section~\ref{sec:conclusions}. 
Throughout the paper, we adopt a \citet{chab03} initial mass function and cosmological parameters of $H_0=67.7$ km s$^{-1}$ Mpc$^{-1}$, $\Omega_\Lambda=0.689$, and $\Omega_{\rm{m}}=0.310$ based on the {\sl Planck} 2018 results \citep{plan20d}.

\section{Data}\label{sec:data}
\subsection{Targets}\label{sec:targets}

The targets analyzed in this paper were drawn from two ALMA programs in Cycles 8 and 10 (project IDs: 2021.1.00027.S and 2023.1.01090.S, respectively; Table~\ref{tab:observations}). 
The Cycle 8 program observed the CO($J=2$--1) line in five well-localized, spectroscopically confirmed FRB hosts \citep{hein20a} that were observable with ALMA and had not previously been observed in CO. 
This paper presents the previously unpublished observations of the hosts of FRBs 20190611B, 20190714A, and 20200430A; the observations of FRBs 20190711A and 20191001A were reported by \citet{hats22} and \citet{yama24a}, respectively. 
The hosts of FRBs 20190714A and 20200430A were also observed in CO(3--2) by \citet{chit23}. 

In Cycle 10, all known accurately localized and spectroscopically confirmed FRB hosts at $z < 0.2$ \citep{hein20a,gord23} that were well observable with ALMA at the time of target selection (April 2023) were selected. 
This redshift limit was adopted 
(i) to minimize possible effects of redshift evolution in host-galaxy properties, 
(ii) to enable a fair comparison with samples of nearby galaxies, and 
(iii) to obtain secure constraints on CO luminosities. 
For this Cycle 10 sample, the CO(1--0) line was observed to derive molecular gas masses without introducing uncertainties associated with CO line luminosity ratios. 
The resulting sample analyzed here comprises nine FRB hosts at $z=0.05$--0.38. 
FRB\,20201124A is a repeating source, while the other eight are treated here as apparent non-repeaters. 

\begin{figure}
\includegraphics[width=\linewidth]{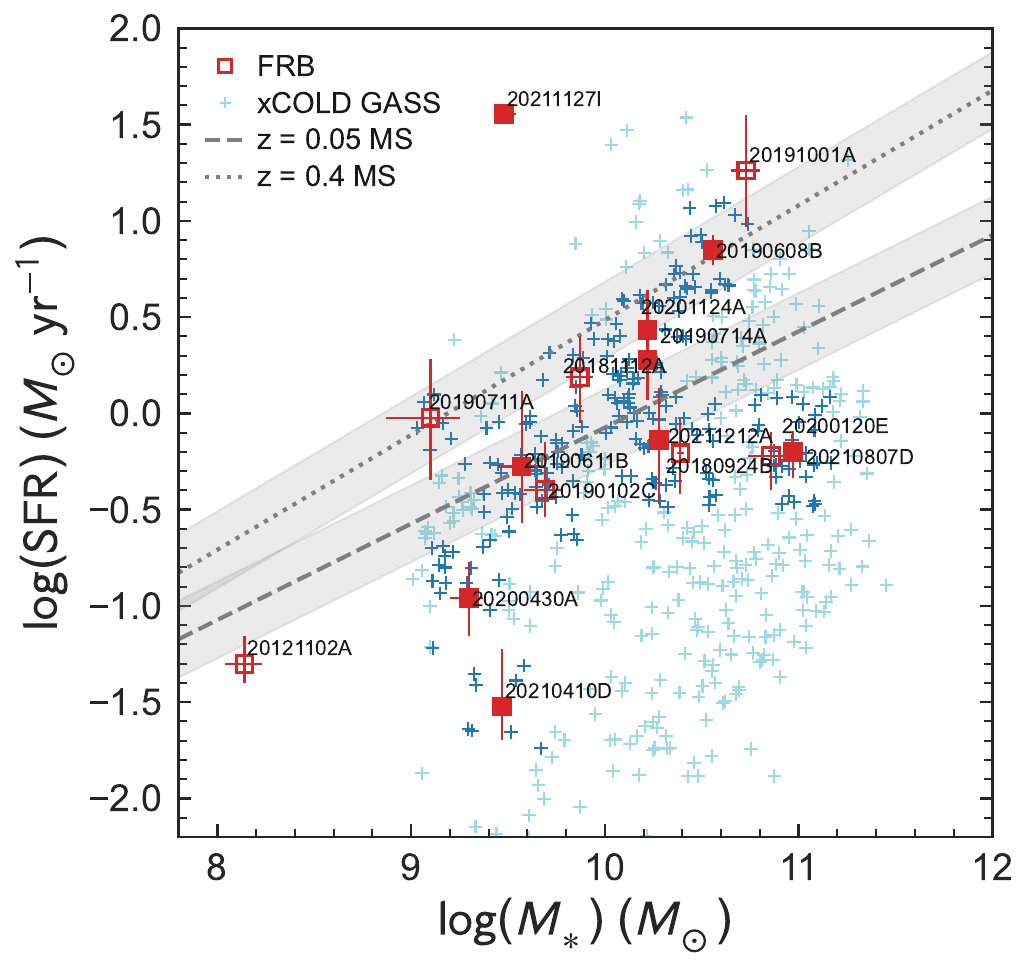}
\caption{
Stellar mass and SFR of FRB hosts with molecular gas observations included in this study. 
The FRB hosts from our new observations are presented as filled red squares.
The stellar mass and SFR are taken from \citet{gord23}. 
For comparison, the local galaxy sample of xCOLD GASS \citep{sain17} is plotted as blue crosses.
The darker, thicker blue crosses indicate the stellar-mass- and SFR-matched control sample used in Section~\ref{sec:properties}. 
The dashed and dotted gray lines show the main sequence of star-forming galaxies at 
$z=0.05$ and $z=0.4$, respectively \citep{spea14}. The shaded regions indicate a representative 0.2 dex interval around each relation.
}
\label{fig:mstar-sfr}
\end{figure}

The stellar masses and SFRs of the nine FRB host targets are plotted in Figure~\ref{fig:mstar-sfr}, together with six FRB hosts from previous molecular-gas studies that are included in the analysis \citep{bowe18, hats22, chit23, yama24a}. 
The targets span wide ranges in stellar mass ($\log{(M_*/M_\odot)} = 9$--11) and SFR (0.03--36 $M_\odot$~yr$^{-1}$). 
The reported gas-phase metallicity estimates span $12+\log({\rm O/H})=8.43$--9.03 \citep{hein20a,gord23}. 
The hosts of FRBs\,20210807D and 20211127I have values of 8.43 and 8.44, respectively, below the solar abundance of $12+\log({\rm O/H})_\odot=8.69$ \citep{aspl21}. 
The remaining seven targets have reported values of 8.72--9.03 and are approximately solar or supersolar. 

For comparison, we plot a local galaxy sample and the MS of star-forming galaxies at $z=0.05$ and $z=0.4$; these redshifts bracket the range spanned by the FRB-host sample.
The majority of FRB hosts, including our targets, are distributed along or near the MS, whereas some deviate significantly from it.

\subsection{Observations and Data Reduction}

The CO(2--1) observations of the hosts of FRBs 20190611B, 20190714A, and 20200430A were conducted in Band 5 during Cycle 8. 
Four basebands with a bandwidth of 1875 MHz (488.28 kHz $\times$ 3840 channels) were used in the 2$\times$2-bit spectral mode. 
The CO(1--0) observations of the hosts of FRBs 20190608B, 20201124A, 20210410D, 20210807D, 20211127I, and 20211212A were conducted in Band 3 \citep{clau08} during Cycle 10. 
Four basebands were used: one in the 4$\times$4-bit spectral mode with a bandwidth of 937.5 MHz to improve the line sensitivity for the CO emission line, and the remaining three in the standard 2$\times$2-bit spectral mode with a bandwidth of 1875 MHz. 
The ALMA observations are summarized in Table~\ref{tab:observations}. 

The data were reduced with Common Astronomy Software Applications (CASA; \cite{mcmu07,the22}). 
Data calibration was performed using the ALMA Science Pipeline in CASA. 
Image cubes and continuum maps were produced with the \verb|tclean| task using Briggs weighting. 
The CO maps were created with a \verb|robust| parameter of 2.0 for the hosts of FRBs 20190611B, 20190714A, 20200430A, 20210410D, and 20210807D to improve surface-brightness sensitivity, and 0.5 for the remaining hosts. 
All continuum maps were created with a \verb|robust| parameter of 2.0. 
CLEAN masks were placed for imaging CO and continuum emission when components with peak signal-to-noise ratios (S/N) above five were identified. 
The continuum maps were formed from line-free channels by excluding channels containing CO emission. 

\begin{table*}
\tbl{ALMA Observations}{
\begin{tabular}{cccccccccccc}
\hline
FRB & $z$ & Scale & CO & $\nu_{\rm obs}$ & Band & $T_{\rm on}$ & $N_{\rm ant}$ & Baseline & Beam size & $\sigma_{\rm line}$ & $\sigma_{\rm cont}$\\
 & & (kpc arcsec$^{-1}$) & & (GHz) & & (min) & & (m) & (arcsec) & (mJy beam$^{-1}$) & ($\mu$Jy beam$^{-1}$) \\
(1) & (2) & (3) & (4) & (5) & (6) & (7) & (8) & (9) & (10) & (11) & (12) \\
\hline
\multicolumn{12}{c}{Cycle 8} \\
\hline
20190611B 
& 0.3778 & 5.35 & 2--1 & 167.3233 & 5 & 316 & 43--46 & 15.1--1301  & $1.01\times 0.75$ & 0.15 & 6.3\\
20190714A$^\dagger$
& 0.2365 & 3.87 & 2--1 & 186.4440 & 5 &  46 & 46    & 15.1--783.5 & $1.05\times 0.60$ & 1.1  & 18\\
20200430A$^\dagger$
& 0.1608 & 2.86 & 2--1 & 198.6027 & 5 &  24 & 44    & 15.1--783.5 & $0.80\times 0.56$ & 0.52 & 22\\
\hline
\multicolumn{12}{c}{Cycle 10} \\
\hline
20190608B                
& 0.1178 & 2.20 & 1--0 & 103.1251 & 3 & 102 & 46    & 15.1--483.9 & $1.82\times 1.66$ & 0.30 & 8.3\\
20201124A
& 0.0979 & 1.87 & 1--0 & 104.9924 & 3 &  62 & 43    & 15.1--783.5 & $1.29\times 1.01$ & 0.43 & 11\\
20210410D 
& 0.1415 & 2.57 & 1--0 & 100.9822 & 3 & 264 & 43    & 15.1--2046  & $1.24\times 0.86$ & 0.17 & 4.7\\
20210807D 
& 0.1293 & 2.38 & 1--0 & 102.0731 & 3 & 114 & 42--45 & 15.1--1604  & $1.07\times 0.90$ & 0.25 & 6.7\\
20211127I 
& 0.0469 & 0.95 & 1--0 & 110.1072 & 3 &   5 & 46    & 15.1--483.9 & $2.03\times 1.51$ & 1.2 & 29\\
20211212A 
& 0.0707 & 1.39 & 1--0 & 107.6597 & 3 &  14 & 43    & 15.1--649.6 & $1.47\times 1.29$ & 0.65 & 21\\
\hline
\end{tabular}}\label{tab:observations}
\begin{tabnote}
Columns 
(1) FRB name; 
(2) Redshift; 
(3) Angular scale calculated at the spectroscopic redshift using the cosmological parameters adopted in this paper; 
(4) CO rotational transition; 
(5) Redshifted frequency of the CO line; 
(6) ALMA Band; 
(7) On-source integration time; 
(8) Number of antennas; 
(9) Range of baseline lengths; 
(10) Synthesized beam size; 
(11) RMS noise level at a velocity resolution of 10 km~s$^{-1}$; 
(12) RMS noise level for continuum. \\
$^\dagger$ CO(3--2) observations of these hosts were reported by \citet{chit23}. \\
\end{tabnote}
\end{table*}

\section{Results}\label{sec:results}

CO emission was significantly detected (S/N $>$ 5) in six hosts (FRBs 20190608B, 20190714A, 20201124A, 20210807D, 20211127I, and 20211212A) and tentatively detected (S/N $\sim$ 3.9) in the FRB\,20210410D host. 
This result doubles the number of FRB hosts in which CO has been detected. 
CO is not significantly detected in the hosts of FRBs 20190611B and 20200430A. 
Continuum emission is tentatively detected in three hosts: FRB 20190608B (peak S/N = 3.7), FRB 20201124A (peak S/N = 4.1), and FRB 20210410D (peak S/N = 3.7).
The CO velocity-integrated intensity and millimeter continuum maps are shown in Figure~\ref{fig:map}. 
The CO emission in the six hosts is spatially resolved by the synthesized beam, and the CO fluxes are measured by integrating emission above 2$\sigma$. 
Among the hosts identified as spiral galaxies in the optical images, the CO emission in the FRB\,20190608B host is concentrated toward the central region, whereas it extends into the outer regions of the FRB\,20210807D and FRB\,20211127I hosts. 
In the FRB\,20211212A host, the detected CO emission extends beyond the optical emission. 
The CO surface brightness at the FRB position is low relative to the brighter host emission for FRBs 20190608B, 20190714A, 20210410D, and 20210807D, showing that the FRB positions do not necessarily coincide with the brightest molecular-gas regions on the spatial scales probed here. 
The three FRB hosts where CO was not significantly detected in our ALMA observations (FRBs 20190611B, 20200430A, and 20210410D) have relatively low stellar masses ($\log{M_*}<9.6$) and lie at the higher-redshift end of our sample ($z>0.14$). 

The CO intensity-weighted velocity fields for the six CO-detected hosts are shown in Figure~\ref{fig:mom1}. 
The maps reveal resolved, large-scale velocity gradients with projected amplitudes of order 100--200 km~s$^{-1}$ across the CO-emitting regions. 
The hosts of FRBs 20201124A, 20211127I, and 20211212A exhibit relatively smooth velocity gradients consistent with ordered disk rotation at the spatial resolution of the observations. 
The centrally concentrated CO emission in the FRB\,20190608B host also shows a systematic velocity gradient, although its limited spatial extent prevents a detailed characterization of the gas kinematics. 
We find no clear velocity discontinuity or distinct kinematic component associated with the FRB positions. 
However, the current kpc-scale resolution and limited number of independent synthesized beams do not exclude non-circular or disturbed motions on smaller spatial scales. 

Figure~\ref{fig:spectra} shows the CO spectra. 
Single-Gaussian fits yield full widths at half maximum (FWHM) of approximately 100--300 km~s$^{-1}$, comparable to those of nearby star-forming galaxies \citep{both14,cico17}. 
The fitted CO redshifts and line widths are listed in Table~\ref{tab:results}.

In the following subsections, we describe the CO and millimeter continuum properties of the individual host galaxies. 

\subsection{FRB\,20190608B}
The host galaxy of FRB\,20190608B is a barred spiral galaxy at $z=0.11778$, and high-resolution optical and integral-field spectroscopic observations show that the FRB position is coincident with a star-forming knot in a spiral arm, with no clear evidence for disturbed host morphology or kinematics \citep{chit21}. 
The host has an SFR of $\sim7.0$ $M_{\odot}$~yr$^{-1}$ \citep{chit21,gord23}. 
Our ALMA CO(1--0) observations detect spatially resolved CO emission with $z_{\rm CO}=0.11792\pm0.00004$ and a line width of $242.1\pm25.1$ km~s$^{-1}$. 
The molecular gas distribution is centrally concentrated compared with the optical spiral structure, and the CO peak is offset from the FRB position in the northwest spiral arm. 
The present kpc-scale CO map shows only weak molecular gas emission at the FRB position. 
The 3.1 mm continuum is tentatively detected (the peak S/N of 3.7) with an integrated flux density of $S_{\rm 3.1mm}=81\pm17$ $\mu$Jy.
Both the CO and continuum emission are stronger in the central region than at the FRB position.

\subsection{FRB\,20190611B}
FRB\,20190611B was detected and localized by the Australian SKA Pathfinder (ASKAP; \cite{hota21}). Its host galaxy was identified in optical follow-up observations and included in early FRB host-galaxy demographic studies \citep{hein20a,bhan22}.
Optical and SED studies estimate a modest host-galaxy SFR of $\sim0.5$ $M_{\odot}$~yr$^{-1}$ \citep{hein20a,bhan22}.
Our ALMA observations do not detect significant CO(2--1) emission and millimeter continuum emission from the host. 

\subsection{FRB\,20190714A}
FRB\,20190714A is an ASKAP-localized burst associated with a star-forming host galaxy, with SFR $\sim0.7$--1.9 $M_{\odot}$~yr$^{-1}$ \citep{hein20a,gord23}. 
Our ALMA CO(2--1) observations detect the host, with $z_{\rm CO}=0.23637\pm0.00012$ and a line width of $306.3\pm70.4$ km~s$^{-1}$. 
The CO emission is detected over the host galaxy but is not strongly enhanced at the FRB position. 
The millimeter continuum map does not show a significant detection.

CO(3--2) emission was previously detected by \citet{chit23}. 
Combining their measurement with our CO(2--1) luminosity gives $L'_{\rm CO(3-2)}/L'_{\rm CO(2-1)}=0.18\pm0.04$, indicating low CO excitation. 

\subsection{FRB\,20200430A}
FRB\,20200430A is another ASKAP-localized event with a spectroscopically confirmed host galaxy \citep{hein20a,bhan22}. 
The host has a low SFR of $\sim0.1$ $M_{\odot}$~yr$^{-1}$ in optical/SED analyses \citep{hein20a,bhan22}. 
Our ALMA CO(2--1) data show no significant CO detection. 
The absence of CO contours over the optical host indicates that the molecular gas surface brightness is low across the region covered by the current observations. 
The continuum map also shows no significant emission.

CO(3--2) was tentatively detected by \citet{chit23}, with $L'_{\rm CO(3-2)}=(0.21\pm0.08)\times10^8$ K~km~s$^{-1}$~pc$^2$. 
Our CO(2--1) nondetection gives a 3$\sigma$ lower limit of $L'_{\rm CO(3-2)}/L'_{\rm CO(2-1)}>0.89$. 
This ratio may indicate highly excited molecular gas despite the low host-integrated SFR, but the tentative CO(3--2) detection and CO(2--1) upper limit prevent a firm conclusion. 
Higher-sensitivity observations of multiple CO transitions are required to constrain the excitation.

\subsection{FRB\,20201124A}
FRB\,20201124A is a nearby repeating FRB in a star-forming host galaxy at $z = 0.0979$ \citep{fong21,piro21a}. 
Milliarcsecond localization places the burst in the host galaxy and enables comparison with local star-forming structures \citep{nimm22}. 
Our ALMA CO(1--0) observations detect bright CO emission with $z_{\rm CO}=0.09790\pm0.00002$ and a line width of $220.7\pm15.8$ km~s$^{-1}$. 
The host has an integrated SFR of $\sim2.7$ $M_{\odot}$~yr$^{-1}$, while radio observations indicate enhanced star formation in the vicinity of the burst \citep{piro21a,gord23}. 
The CO emission is spatially resolved and extends the optical emission of the host, indicating a substantial molecular reservoir associated with the star-forming galaxy. 

We also detect 3.1 mm continuum emission from the host (the peak S/N of 4.1) with an integrated flux density of $S_{\rm 3.1mm}=87\pm16$ $\mu$Jy.
The emission is spatially resolved and exhibits two peaks along the north-south direction, with the northern peak consistent with the FRB location. 
The continuum flux measured at the FRB position is $39\pm13$ $\mu$Jy. 
This spatial distribution is consistent with that seen in 15 GHz observations \citep{brun24}
with the Very Large Array (VLA; \cite{perl11}).
FRB\,20201124A has also been associated with a persistent radio source \citep{piro21a,fong21,ravi22b}.
The origin of the millimeter emission detected in this study is discussed in Section~\ref{sec:prs}.

\subsection{FRB\,20210410D}
The host of FRB\,20210410D is included in the sample of well-localized FRB hosts studied by \citet{gord23}, who classified it as a transitioning galaxy in terms of its position relative to the star-forming main sequence. 
Our ALMA CO(1--0) observations yield a tentative CO detection with S/N $\sim$ 3.9. 
The CO spectrum shows negative feature near $-300$ km~s$^{-1}$. We inspected the spectra rebinned to 20--50 km~s$^{-1}$, but the feature remains visible. The low S/N of the emission limits the significance of the tentative detection.
A single-Gaussian fit gives $z_{\rm CO}=0.14166\pm0.00006$ and a line width of $112.6\pm35.2$ km~s$^{-1}$. 
Consistent with its transitioning classification, the host has a low SFR of $\sim0.03$ $M_{\odot}$~yr$^{-1}$ \citep{gord23}. 
The continuum is also tentatively detected (the peak S/N of 3.7) near the FRB position, with the flux density of $17.2\pm4.7$ $\mu$Jy.

\subsection{FRB\,20210807D}
The FRB\,20210807D host was classified as quiescent by \citet{gord23}, and its inferred star-formation history suggests rejuvenation. 
The host has a current SFR of $\sim0.6$ $M_{\odot}$~yr$^{-1}$ despite evidence for rejuvenation in its star-formation history \citep{gord23}. 
Our ALMA CO(1--0) observations reveal strong CO emission with $z_{\rm CO}=0.12935\pm0.00003$ and a line width of $251.0\pm16.7$ km~s$^{-1}$. 
The molecular gas is spatially extended and reaches beyond the central region of the host, in contrast to the centrally concentrated morphology seen in FRB\,20190608B. 
The CO distribution indicates that a substantial molecular reservoir remains in this optically less active system. 
The CO emission at the FRB position is weak, so the burst does not coincide with the brightest molecular-gas region at the current spatial resolution. 
No significant millimeter continuum emission is detected, consistent with the low current star-forming activity inferred from the host-galaxy properties.

\subsection{FRB\,20211127I}
FRB\,20211127I is hosted by WALLABY J131913--185018, for which \textsc{Hi} emission has been detected in the WALLABY pilot survey \citep{glow23}. 
The host is actively star forming, with SFR $\sim36$ $M_{\odot}$~yr$^{-1}$, and coincides with a 1.4 GHz continuum source \citep{glow23}. 
Our ALMA CO(1--0) observations detect CO emission with $z_{\rm CO}=0.04696\pm0.00001$ and a line width of $113.6\pm7.5$ km~s$^{-1}$. 
The CO emission is spatially extended across the host and broadly follows the optical galaxy. 
Together with the reported \textsc{Hi} detection, this indicates that the host contains both atomic and molecular gas reservoirs. 
The CO morphology shows no obvious compact concentration uniquely associated with the FRB position, and the ALMA continuum map does not reveal significant millimeter continuum emission.

\subsection{FRB\,20211212A}
FRB\,20211212A is part of the sample of accurately localized FRBs whose host-galaxy stellar populations were analyzed by \citet{gord23}. 
The host has an SFR of $\sim0.7$ $M_{\odot}$~yr$^{-1}$ from host-galaxy SED modeling \citep{gord23}. 
Our ALMA CO(1--0) observations detect the host at high significance, yielding $z_{\rm CO}=0.07153\pm0.00001$ and a line width of $108.5\pm7.7$ km~s$^{-1}$. 
The CO-emitting region is more extended than the optical emission, indicating that the molecular gas reaches beyond the region dominated by the stellar light. 
CO emission is present at the FRB location, while its peak lies east of the burst position. 
No significant millimeter continuum emission is detected.

\begin{figure*}
\begin{center}
\includegraphics[height=.24\textheight]{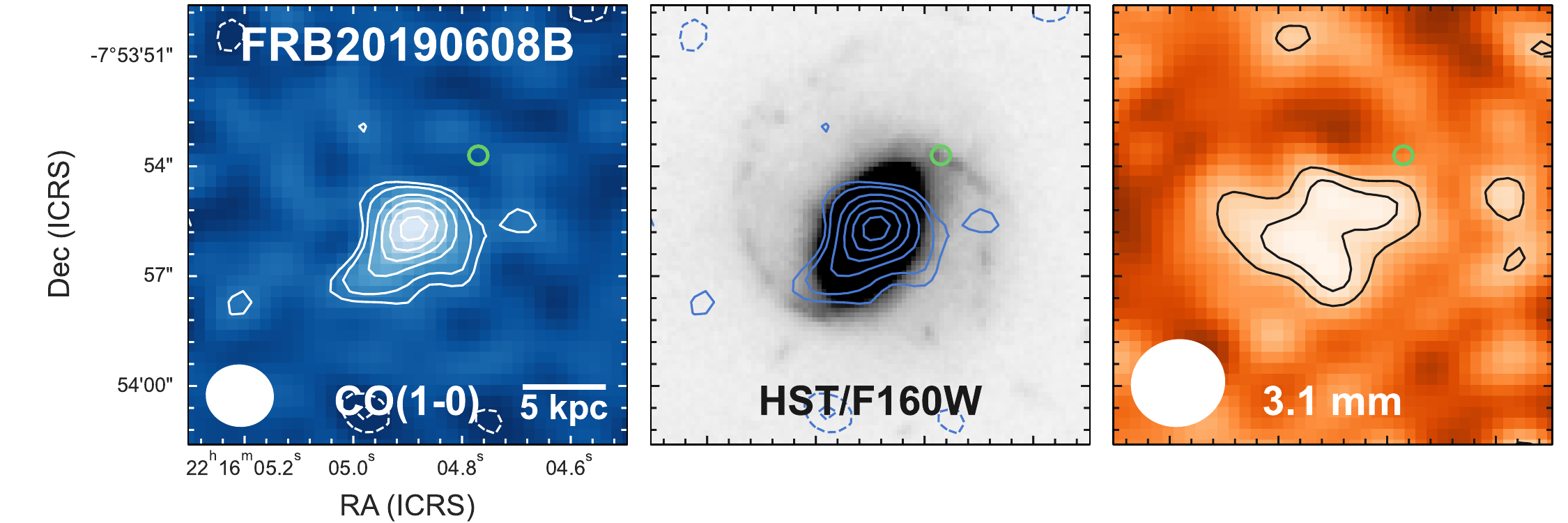}
\par\vspace{1.5ex}
\includegraphics[height=.24\textheight]{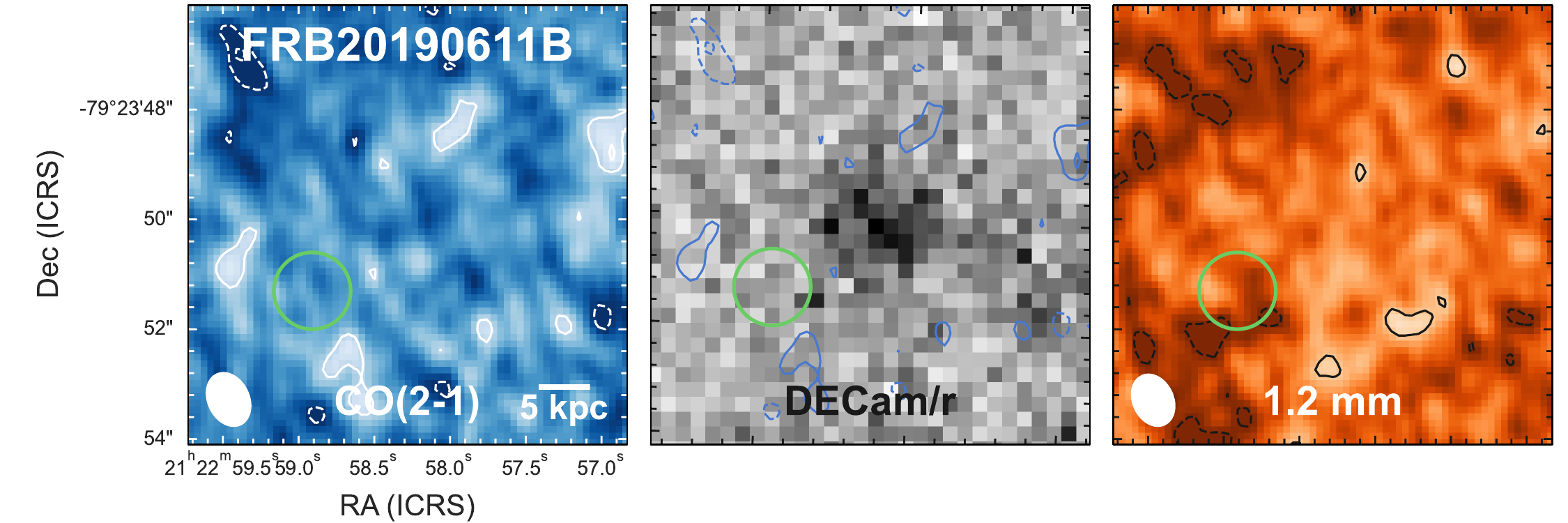}
\par\vspace{1.5ex}
\includegraphics[height=.24\textheight]{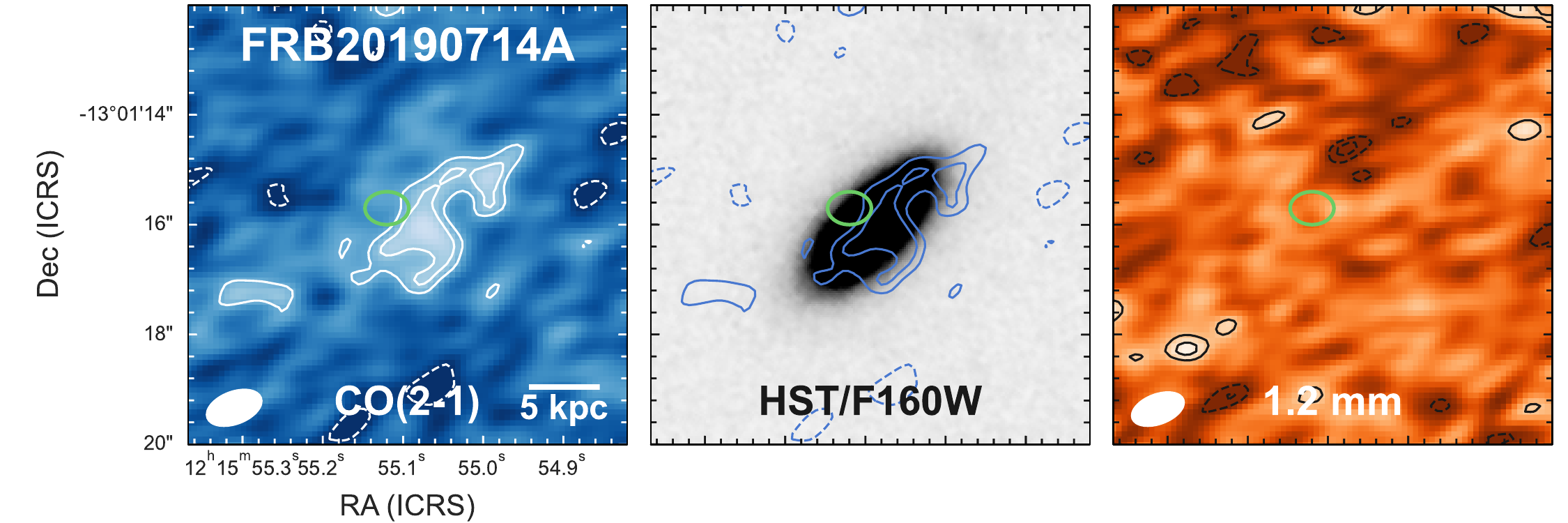}
\end{center}
\caption{
ALMA CO velocity-integrated intensity maps (left), optical images with contours of the CO velocity-integrated intensity (middle), and ALMA continuum maps (right) centered at the host galaxies. 
The contours are 2$\sigma$, 3$\sigma$, and a 2$\sigma$ step subsequently. 
Negative contours are shown with dashed lines. 
The CO maps were created with a \texttt{robust} parameter of 2.0 for the hosts of FRBs 20190611B, 20190714A, 20200430A, 20210410D, and 20210807D to improve surface-brightness sensitivity, and 0.5 for the remaining hosts. 
All continuum maps were created with a \texttt{robust} parameter of 2.0.
The CO maps for hosts with no detection are created by integrating over the channels from $-$200 to $+$200 km~s$^{-1}$. 
The green circle shows the FRB position with uncertainty \citep{hein20a,chit21,gord23,glow23}. 
The FRB\,20201124A is localized with milliarcsecond precision \citep{nimm22}, and only its position is indicated with a cross mark. 
The synthesized beam size is shown in the lower-left corner of the ALMA maps.
}\label{fig:map}
\end{figure*}

\addtocounter{figure}{-1}
\begin{figure*}
\begin{center}
\includegraphics[height=.24\textheight]{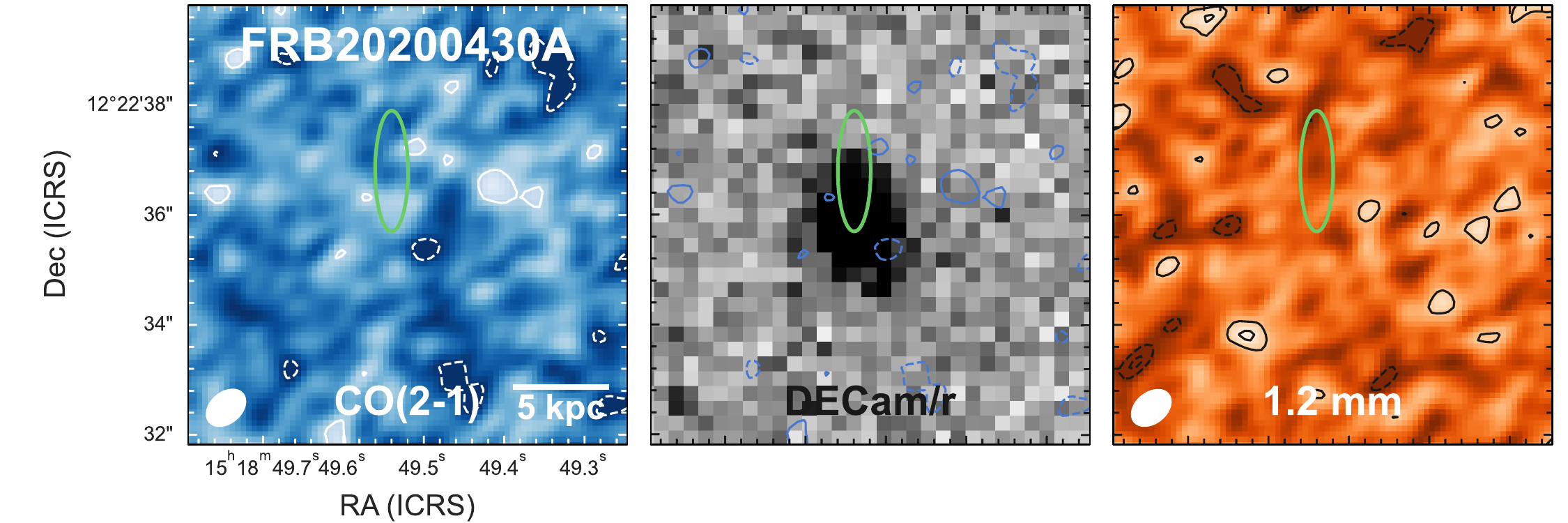}\hspace{.01\linewidth}
\par\vspace{1.5ex}
\includegraphics[height=.24\textheight]{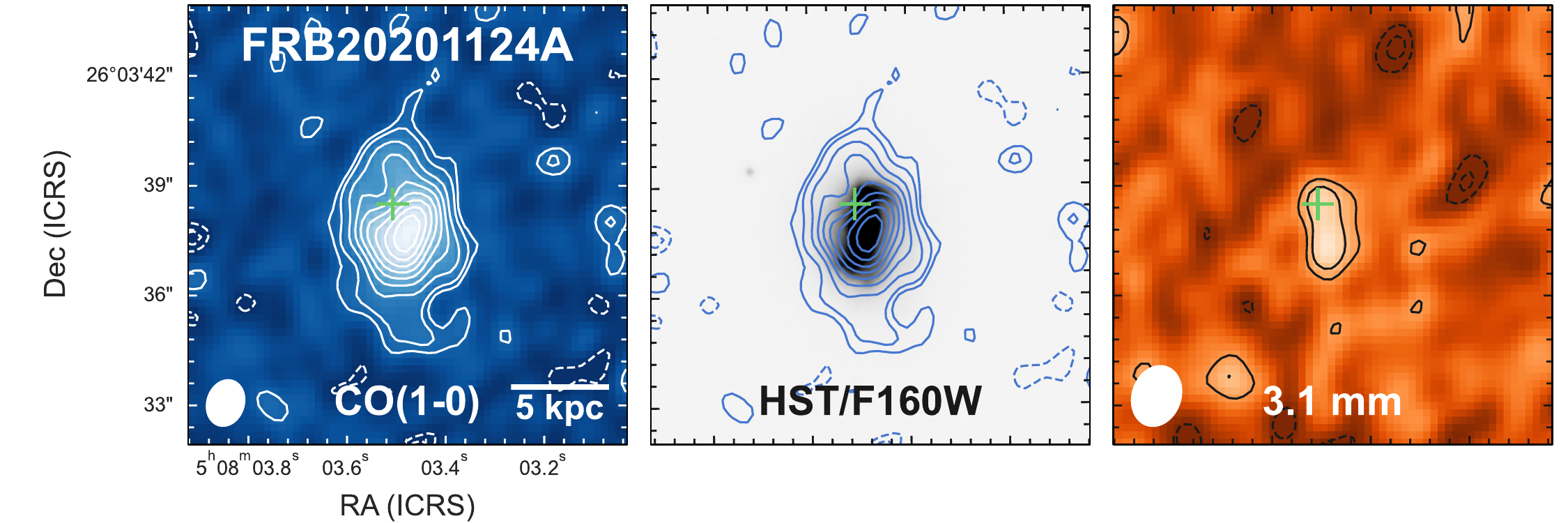}\hspace{.01\linewidth}
\par\vspace{1.5ex}
\includegraphics[height=.24\textheight]{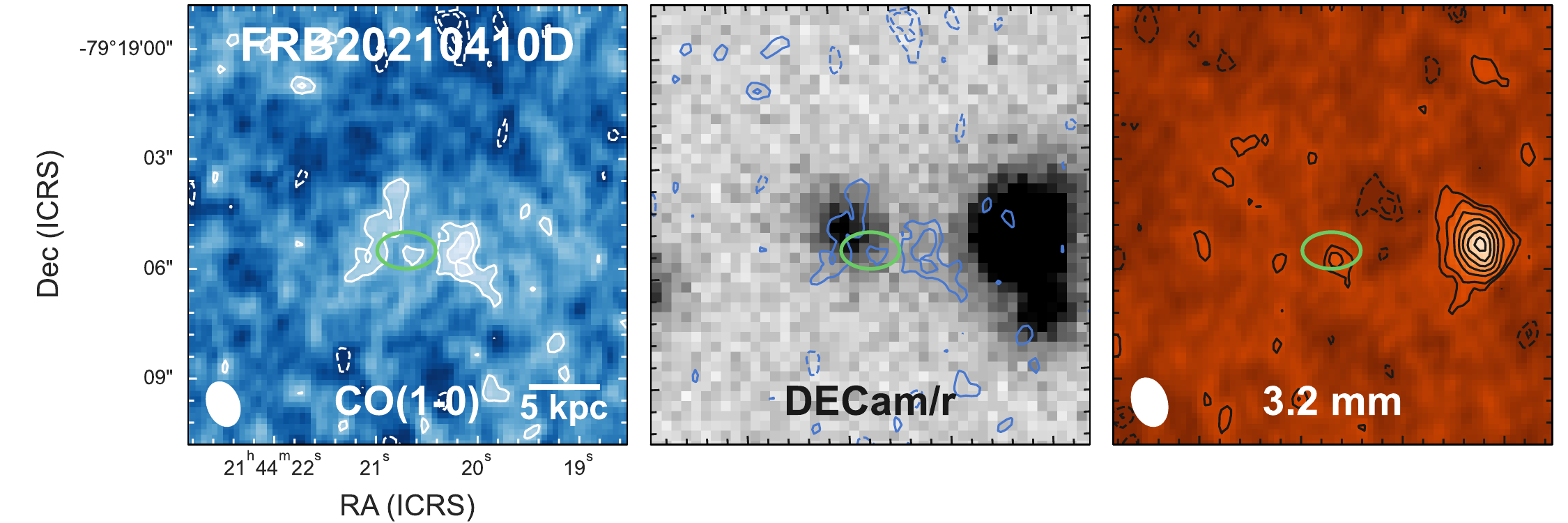}
\end{center}
\caption{Continued.
}
\end{figure*}

\addtocounter{figure}{-1}
\begin{figure*}
\begin{center}
\includegraphics[height=.24\textheight]{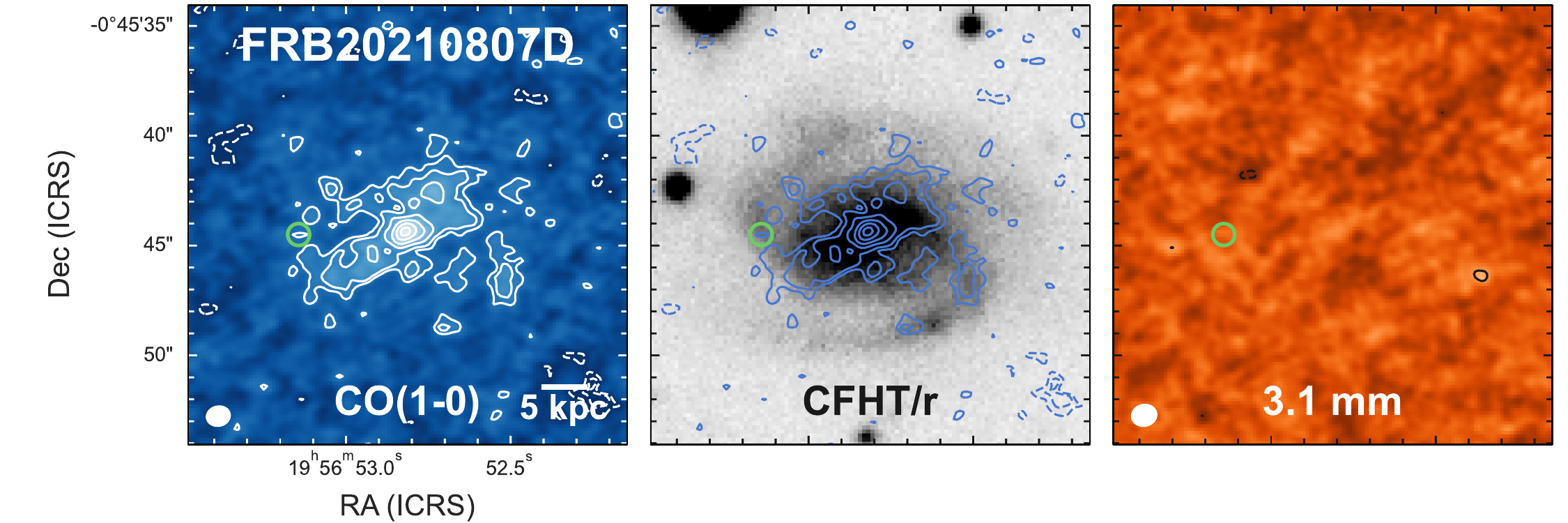}\hspace{.01\linewidth}
\par\vspace{1.5ex}
\includegraphics[height=.24\textheight]{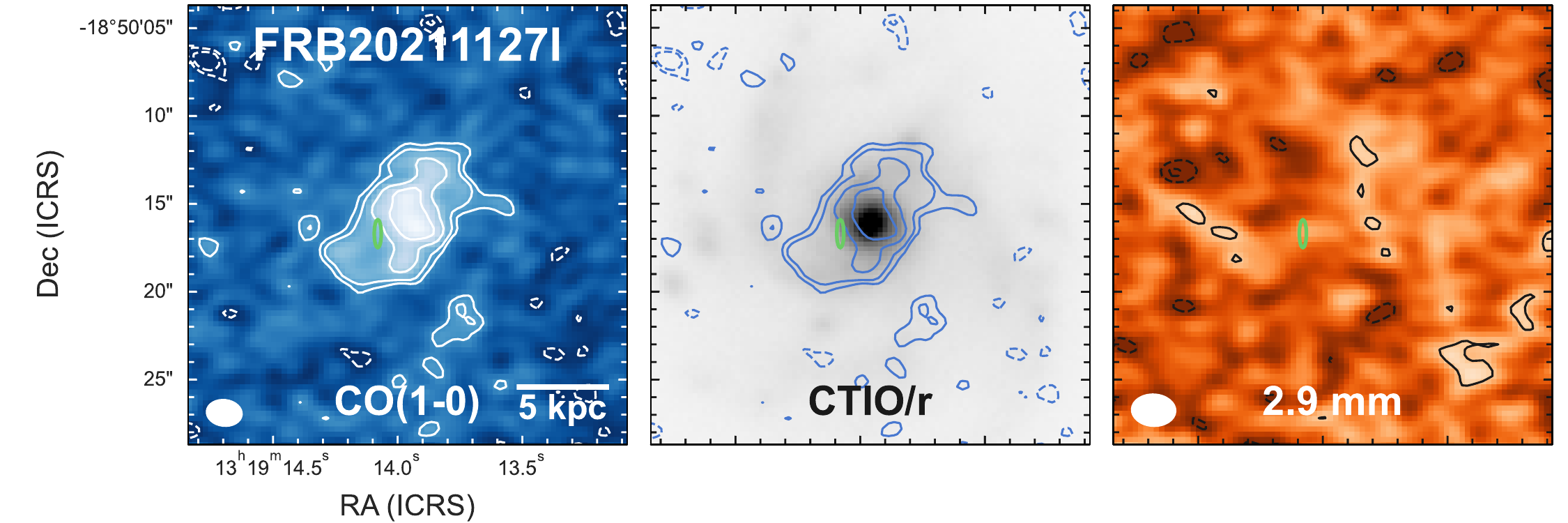}\hspace{.01\linewidth}
\par\vspace{1.5ex}
\includegraphics[height=.24\textheight]{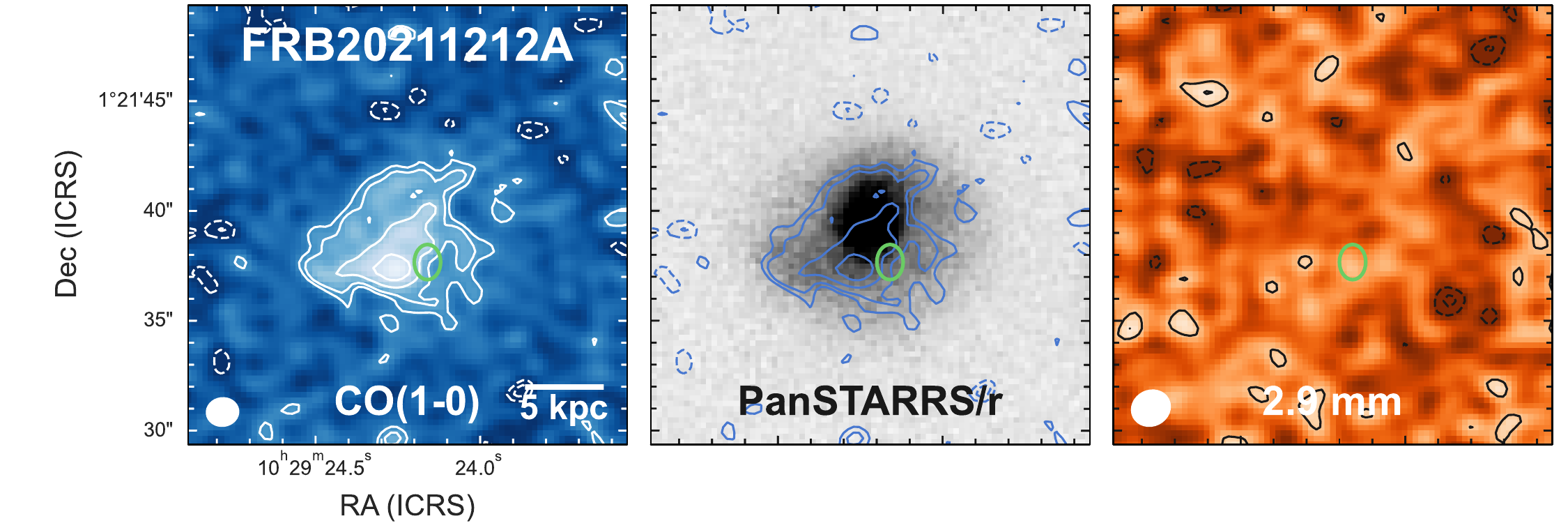}
\end{center}
\caption{Continued.
}
\end{figure*}

\begin{figure*}
\begin{center}
\includegraphics[height=.19\textheight]{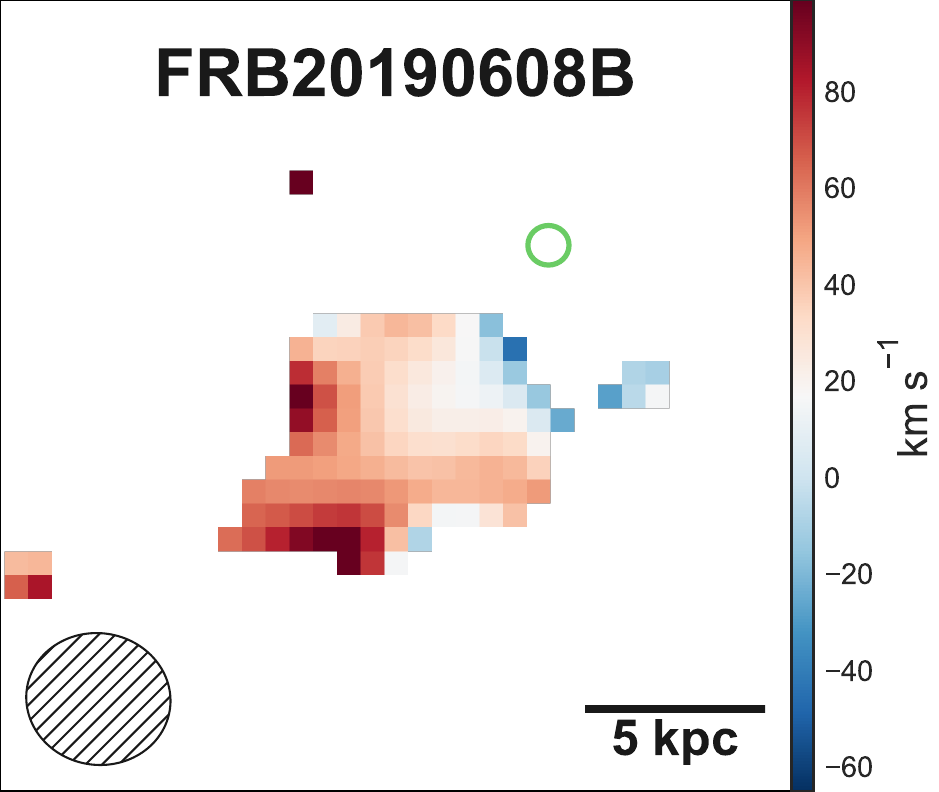}\hspace{.02\linewidth}
\includegraphics[height=.19\textheight]{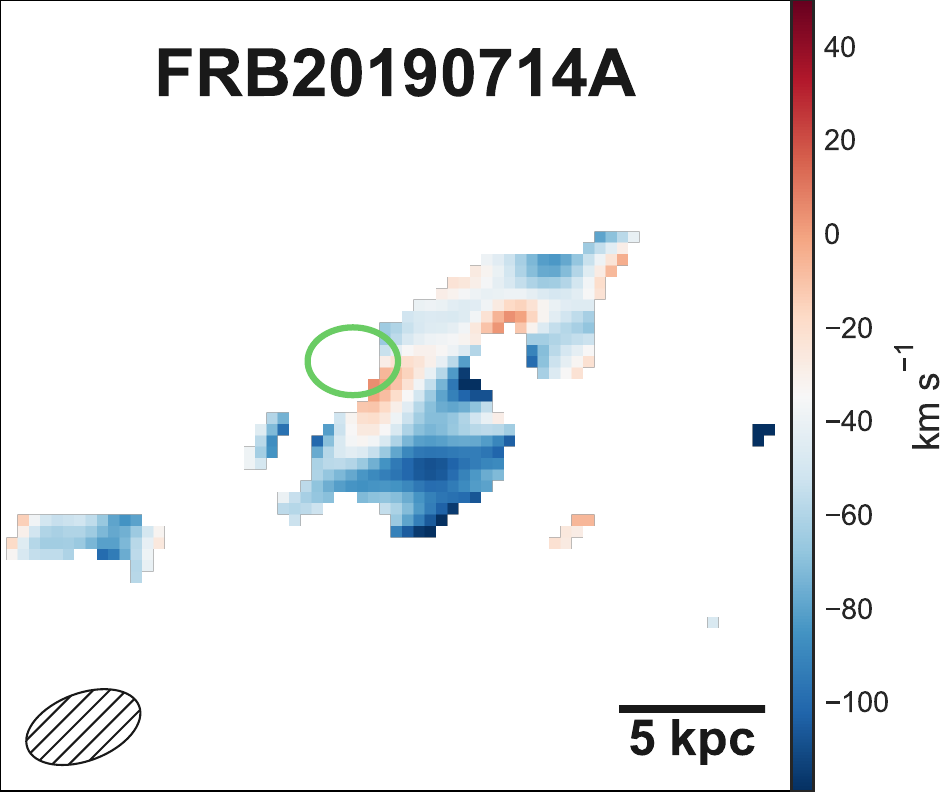}\hspace{.02\linewidth}
\includegraphics[height=.19\textheight]{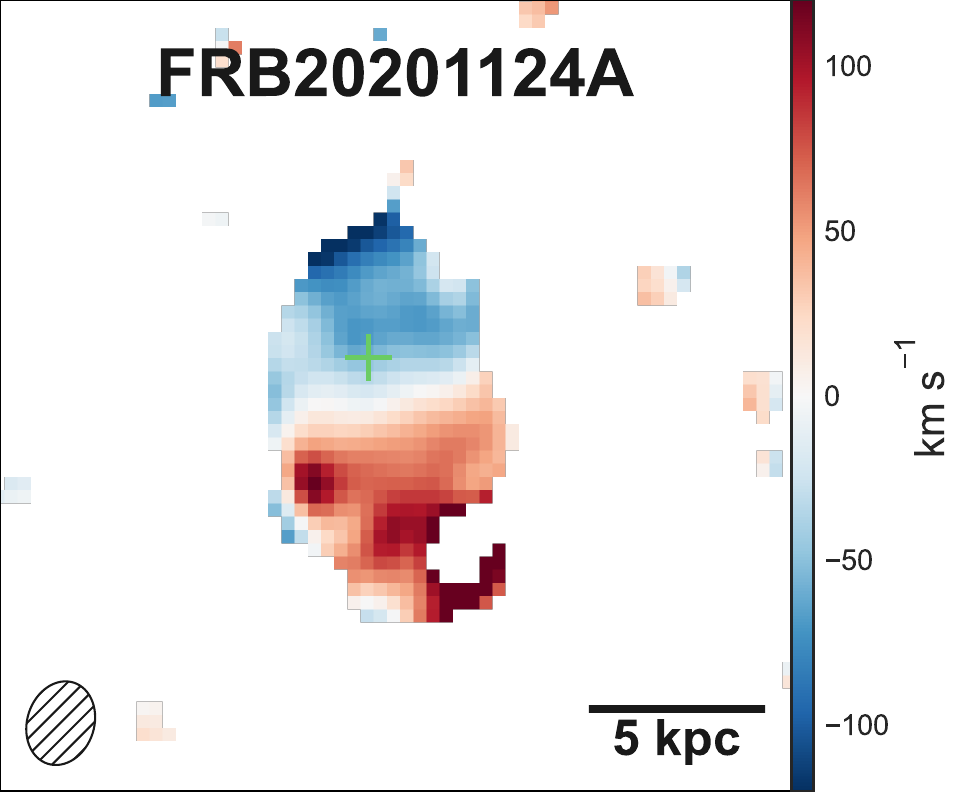}
\par\vspace{2.0ex}
\includegraphics[height=.19\textheight]{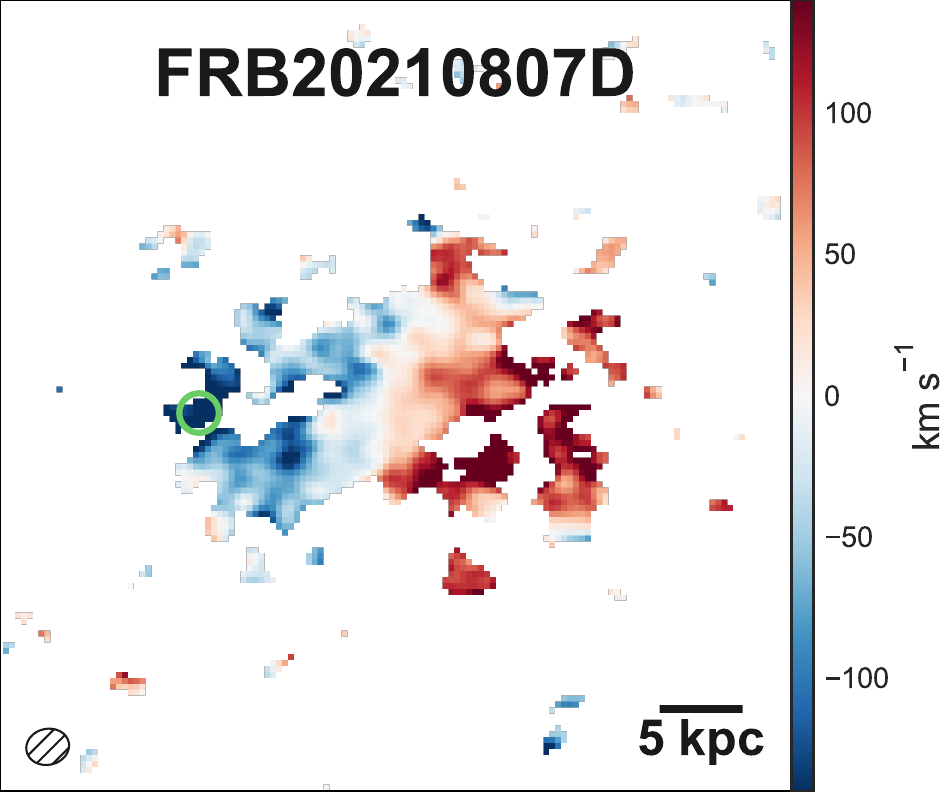}\hspace{.02\linewidth}
\includegraphics[height=.19\textheight]{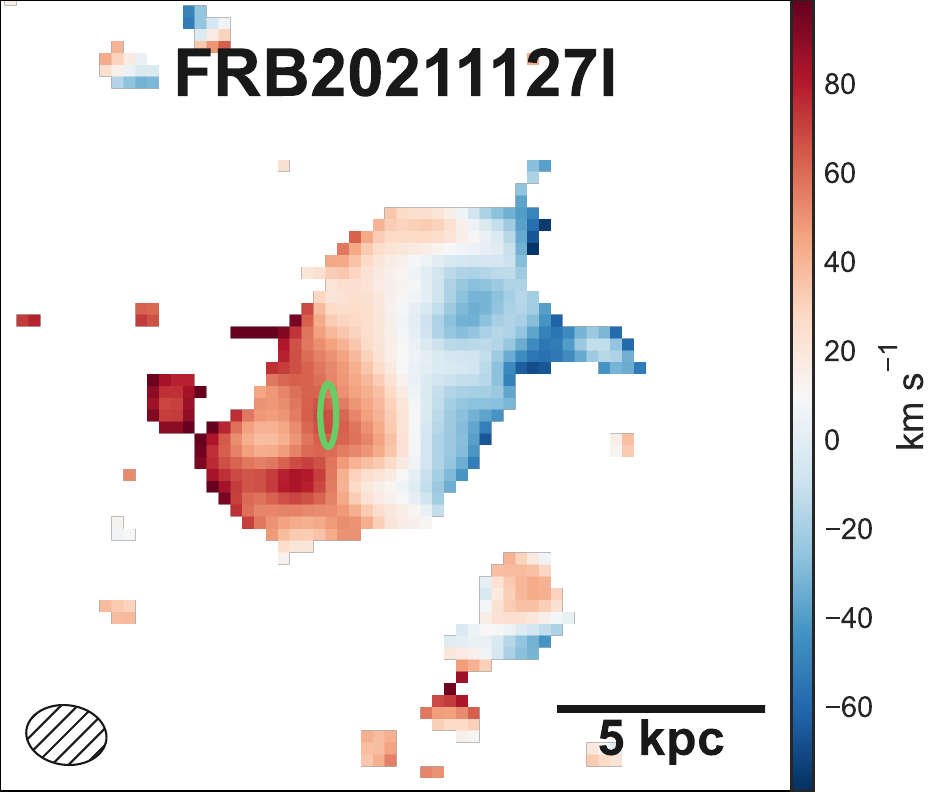}\hspace{.02\linewidth}
\includegraphics[height=.19\textheight]{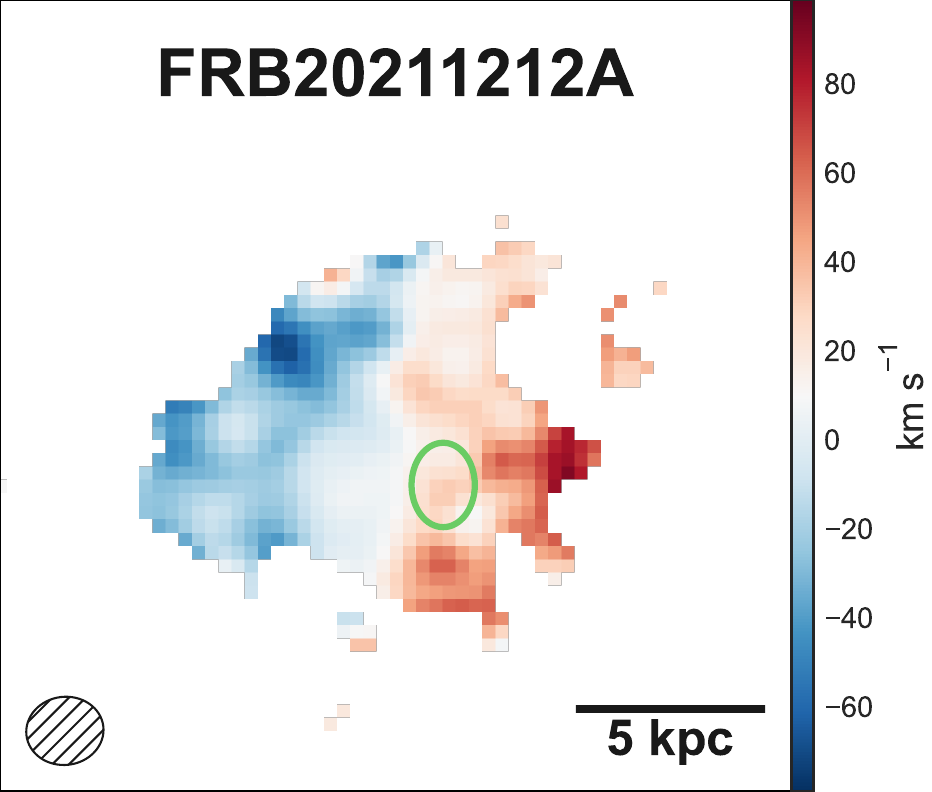}
\end{center}
\caption{
ALMA CO intensity-weighted velocity (moment 1) maps of the FRB host galaxies with detected CO emission. 
Velocities are measured relative to the systemic velocity of each host. 
The green cross shows the FRB position with uncertainty. 
The synthesized beam is shown in the lower-left corner of each panel.
}
\label{fig:mom1}
\end{figure*}

\begin{figure*}
\begin{center}
\includegraphics[width=.32\linewidth]{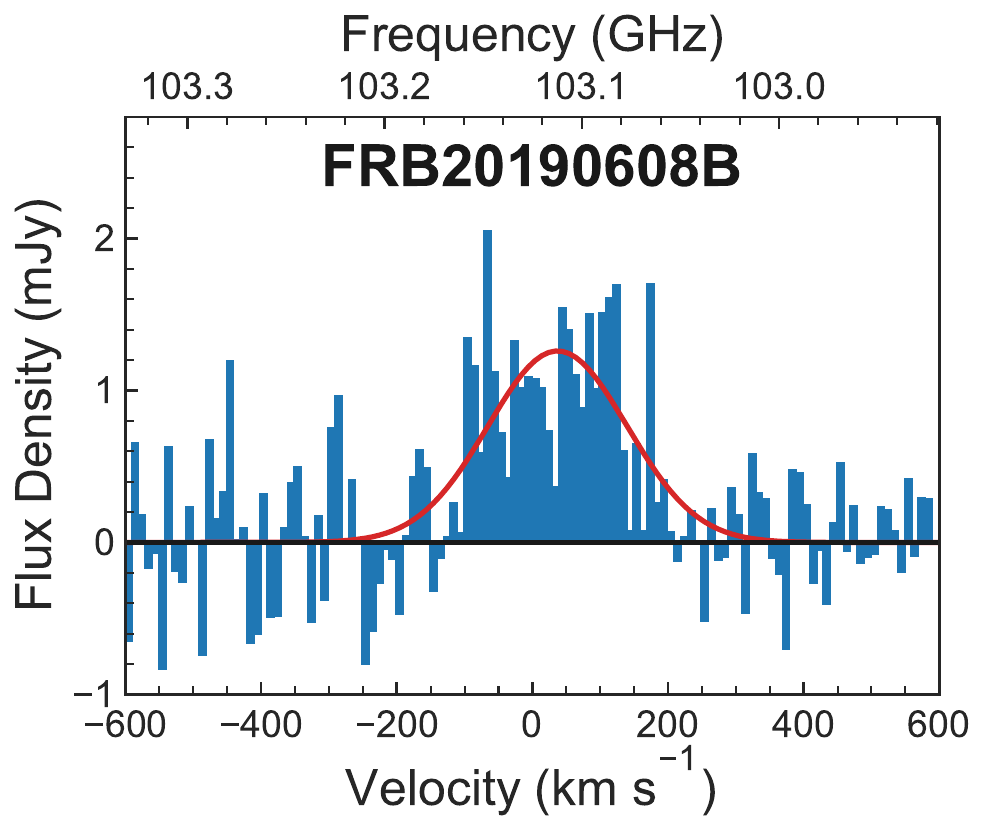}
\includegraphics[width=.32\linewidth]{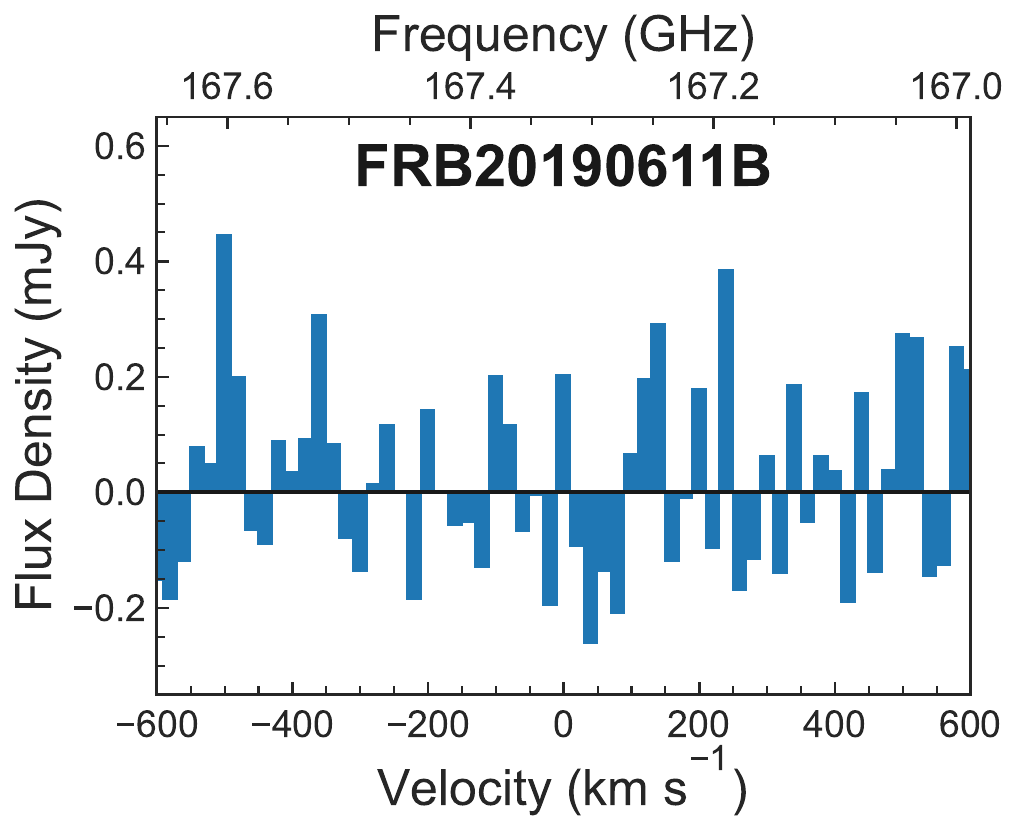}
\includegraphics[width=.32\linewidth]{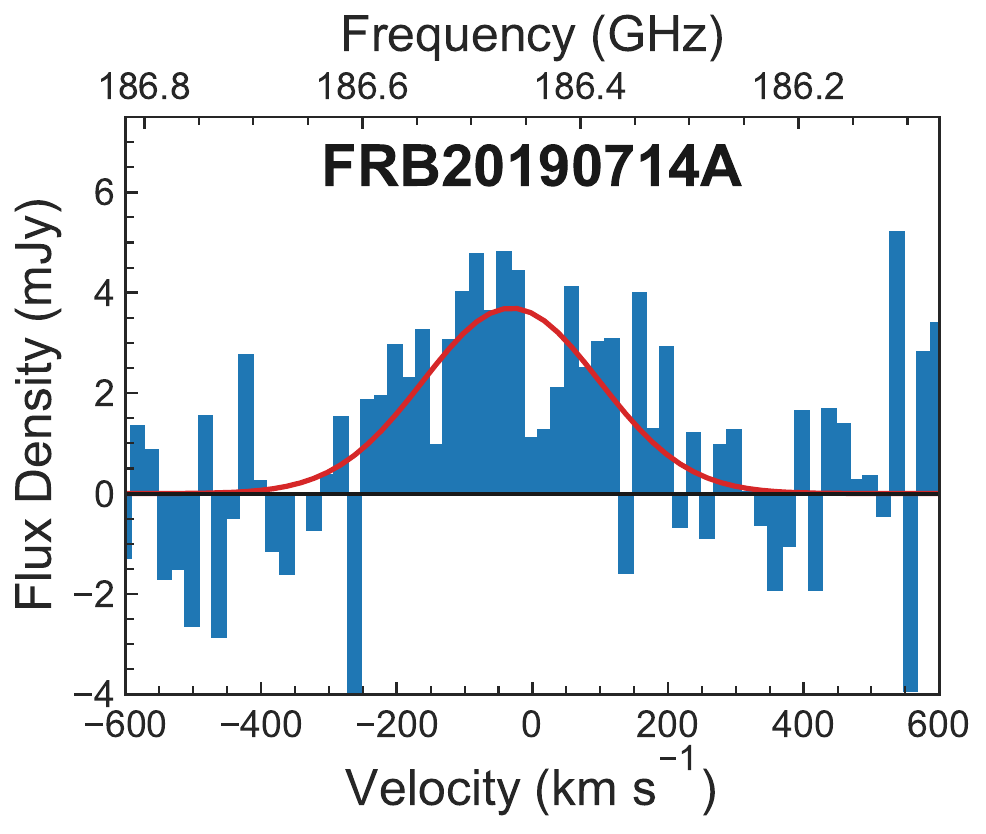}
\par\vspace{1.5ex}
\includegraphics[width=.32\linewidth]{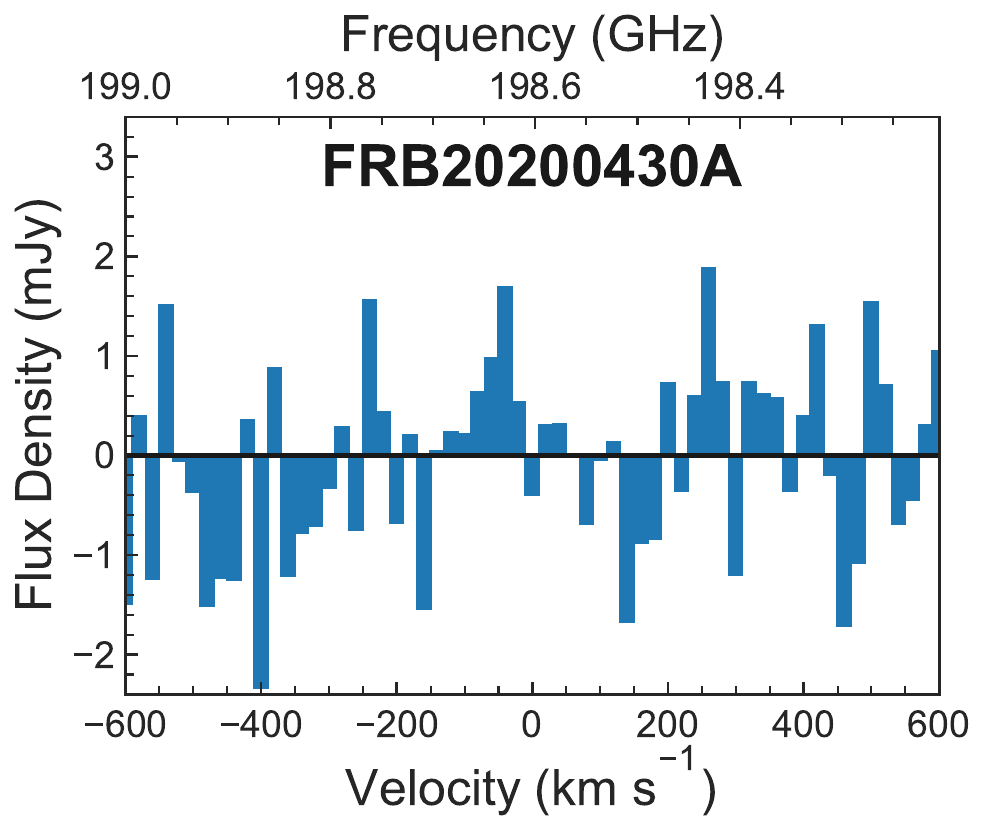}
\includegraphics[width=.32\linewidth]{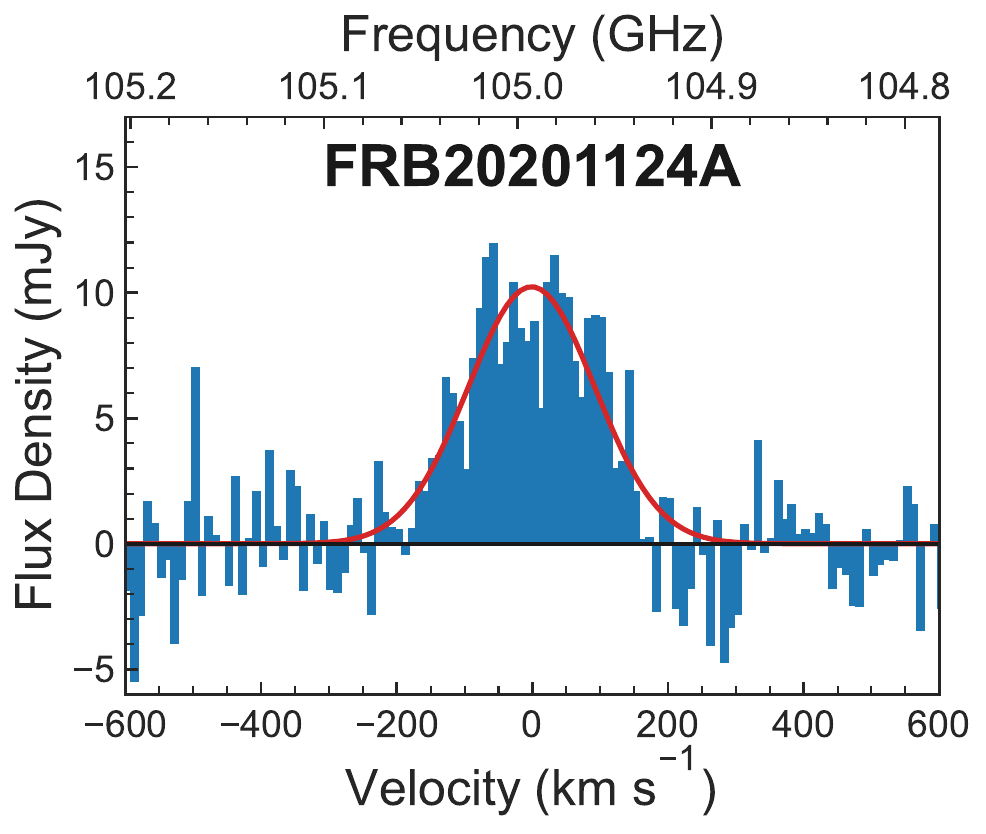}
\includegraphics[width=.32\linewidth]{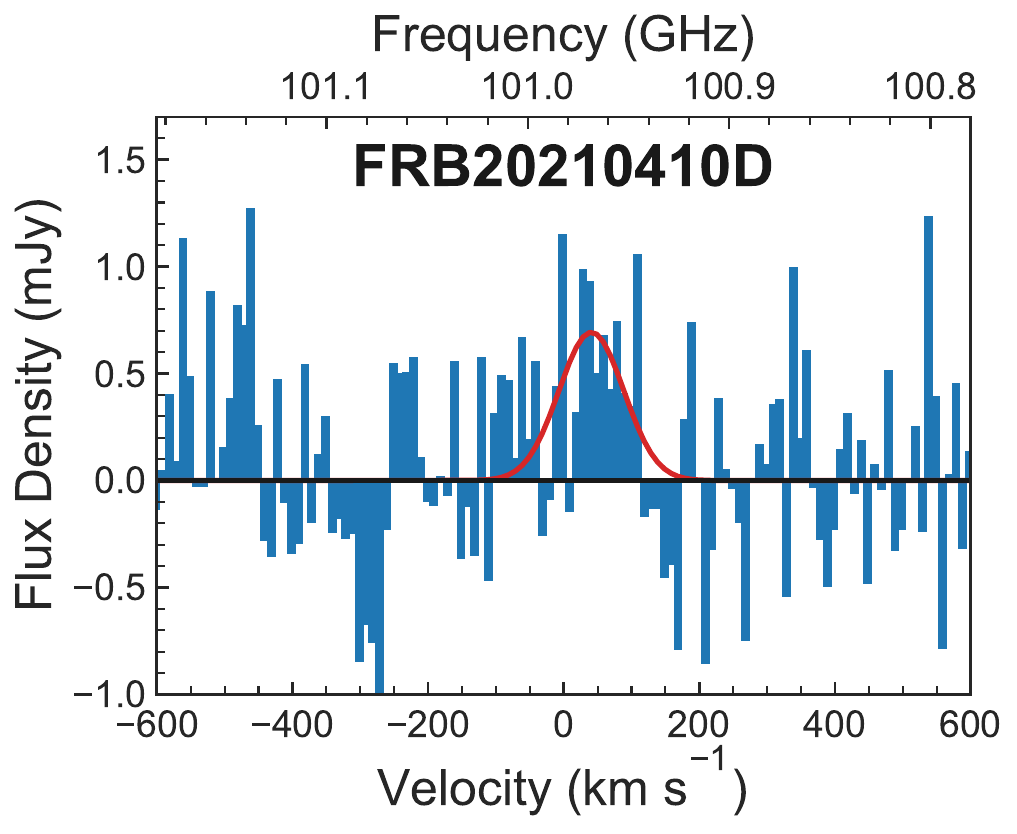}
\par\vspace{1.5ex}
\includegraphics[width=.32\linewidth]{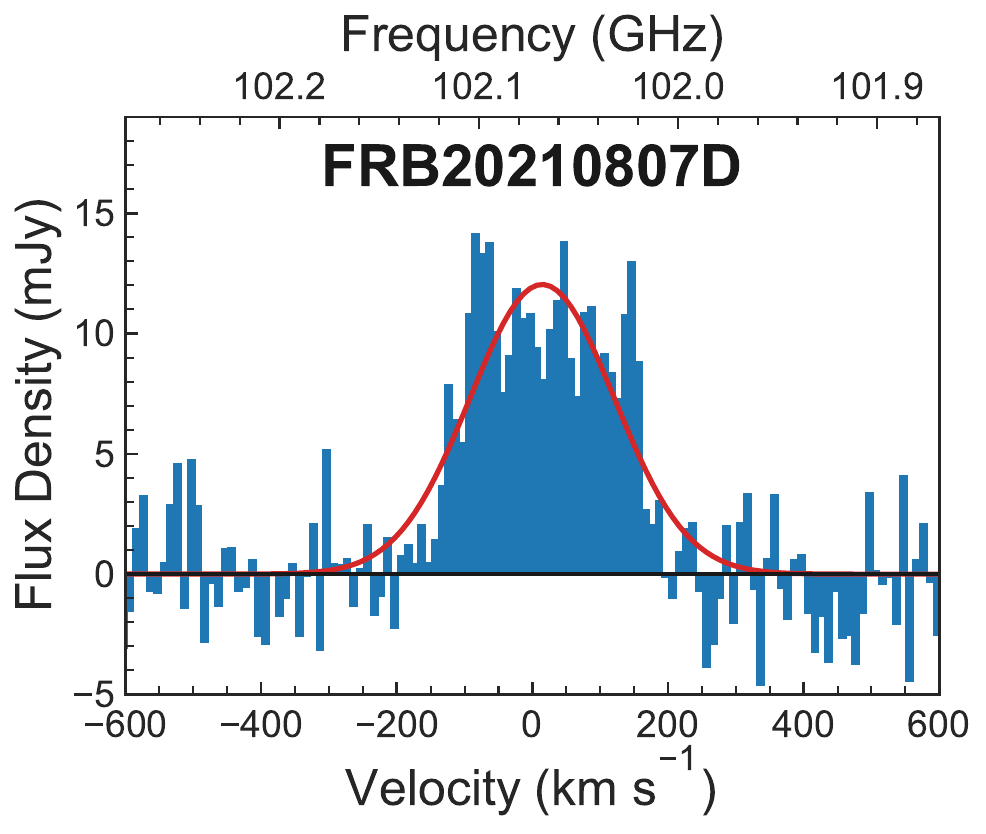}
\includegraphics[width=.32\linewidth]{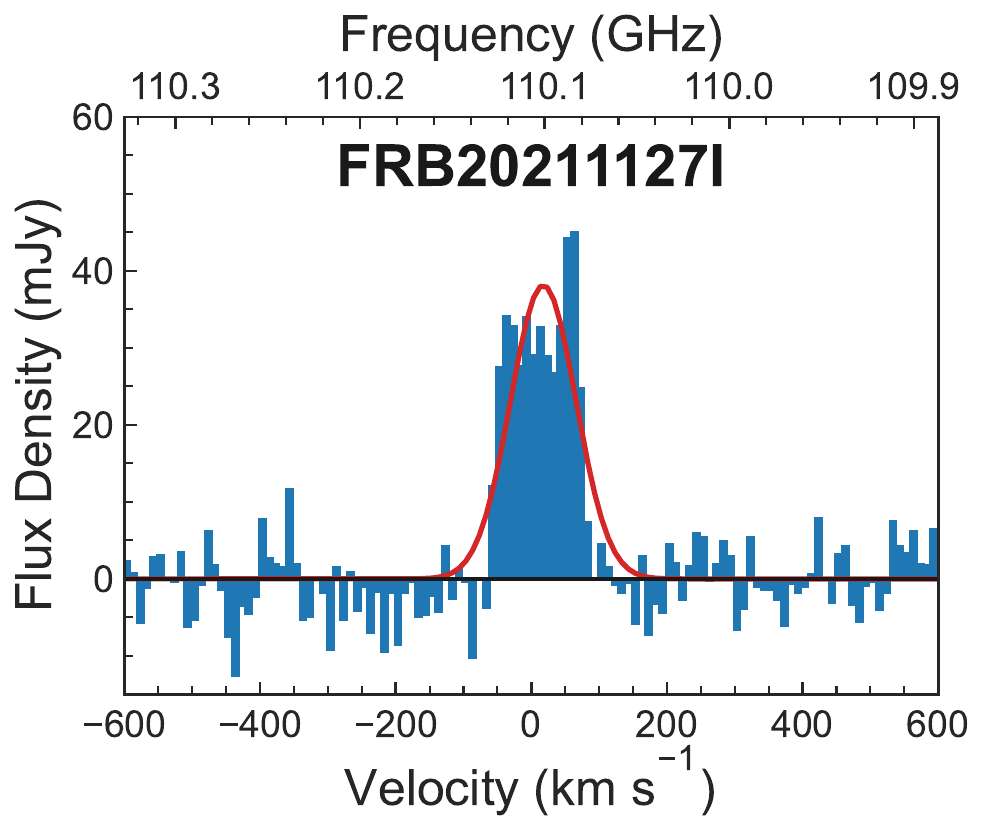}
\includegraphics[width=.32\linewidth]{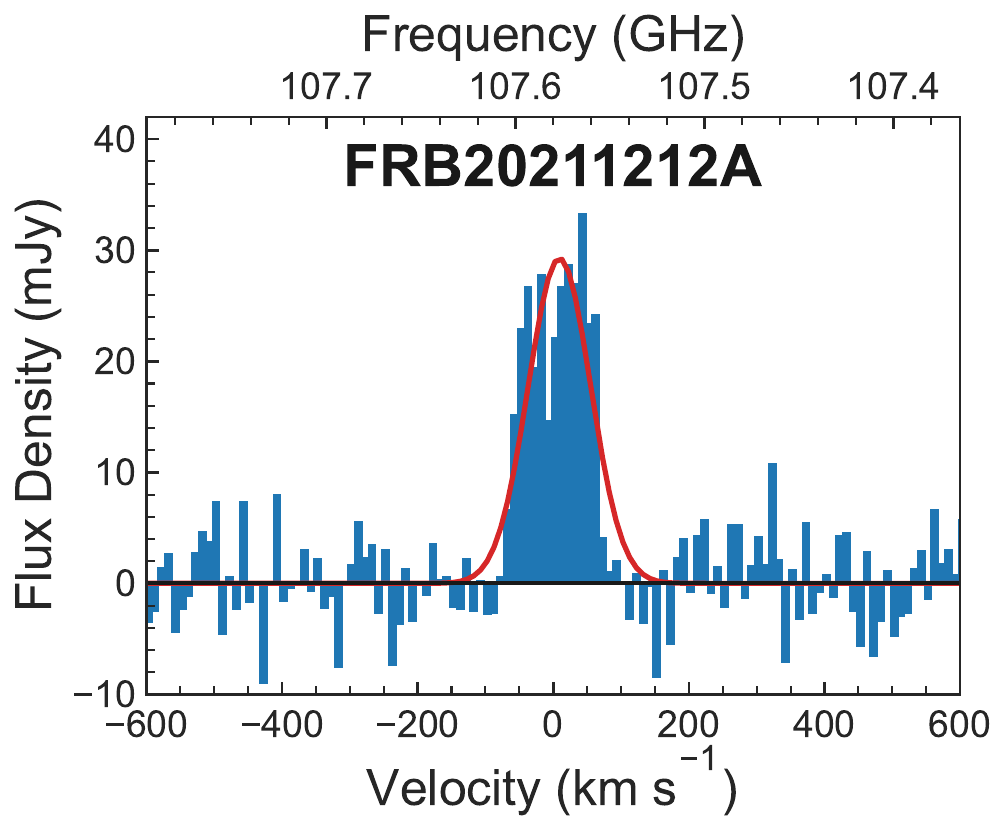}
\end{center}
\caption{CO spectra extracted with apertures encompassing the host galaxies. 
The velocity offset is measured relative to the redshift listed in Table~\ref{tab:observations}.
The velocity resolution is 20 km~s$^{-1}$ for FRBs 20190611B, 20190714A, and 20200430A, and 10 km~s$^{-1}$ for FRBs 20190608B, 20201124A, 20210410D, 20210807D, 20211127I, and 20211212A. 
Red lines show the results of fitting with a single Gaussian.
}\label{fig:spectra}
\end{figure*}

\begin{table*}
\tbl{Results of ALMA Observations}{
\begin{tabular}{ccccccccc}
\hline
FRB & $z_{\rm CO} $ & $\Delta v$ & $S\Delta v$ & S/N & $L^{\prime}_{\rm CO}$ & $M_{\rm mol}$ & $\mu_{\rm mol}$ & $\tau_{\rm gas}$\\
&  & (km s$^{-1}$) & (Jy km s$^{-1}$) & & (K km s$^{-1}$ pc$^2$) & ($M_\odot$) & & (Gyr)\\
(1) & (2) & (3) & (4) & (5) & (6) & (7) & (8) & (9) \\
\hline
20190608B & $0.11792\pm0.00004$ & $242\pm25$ & $0.21\pm0.03$  & 7.7 & $(1.5\pm0.2)\times10^8$ & $(6.3\pm0.8)\times10^8$     & $0.017\pm0.002$ & $0.090\pm0.022$\\
20190611B & --                  & --             & $<0.020$       & -- & $<3.9\times10^7$        & $<2.6\times10^8$            & $<0.069$        & $<0.48$ \\
20190714A & $0.23637\pm0.00012$ & $306\pm70$ & $1.5\pm0.2$    & 6.8 & $(1.1\pm0.2)\times10^9$ & $(7.3\pm1.1)\times10^9$     & $0.44\pm0.08$   & $0.26\pm0.17$ \\
20200430A & --                  & --             & $<0.069$       & -- & $<2.3\times10^7$        & $<1.5\times10^8$            & $<0.076$        & $<1.4$ \\
20201124A & $0.09790\pm0.00002$ & $221\pm16$ & $2.1\pm0.1$    & 24.5 & $(9.9\pm0.4)\times10^8$ & $(4.3\pm0.2)\times10^9$     & $0.26\pm0.03$   & $1.6\pm1.0$ \\
20210410D & $0.14166\pm0.00006$ & $113\pm35$ & $0.049\pm0.013$& 3.9 & $(5.0\pm1.3)\times10^7$ & $(2.2\pm0.6)\times10^8$     & $0.073\pm0.021$ & $7.2\pm7.4$ \\
20210807D & $0.12935\pm0.00003$ & $251\pm17$ & $2.7\pm0.1$    & 26.5 & $(2.3\pm0.1)\times10^9$ & $(1.00\pm0.04)\times10^{10}$& $0.11\pm0.01$   & $16\pm5$ \\
20211127I & $0.04696\pm0.00001$ & $114\pm 8$ & $4.7\pm0.2$    & 20.3 & $(5.1\pm0.3)\times10^8$ & $(2.2\pm0.1)\times10^9$     & $0.73\pm0.11$   & $0.061\pm0.003$\\
20211212A & $0.07153\pm0.00001$ & $109\pm 8$ & $2.7\pm0.1$    & 22.6 & $(6.7\pm0.3)\times10^8$ & $(2.9\pm0.1)\times10^9$     & $0.15\pm0.02$   & $3.9\pm3.3$ \\
\hline
\end{tabular}}\label{tab:results}
\begin{tabnote}
Limits are 3$\sigma$. \\
Columns
(1) FRB name; 
(2) Redshift derived from the CO line for CO-detected sources; 
(3) CO line width (FWHM) obtained by fitting with a single Gaussian; 
(4) Velocity-integrated CO flux. A velocity width of 160 km s$^{-1}$ is assumed for nondetection; 
(5) S/N ratios for the velocity-integrated CO flux;
(6) CO line luminosity of the observed transition;
(7) Molecular gas mass derived with the Milky Way conversion factor, $\alpha_{\rm CO} = 4.3$ $M_{\odot}$ (K km~s$^{-1}$ pc$^2$)$^{-1}$;
(8) Molecular gas fraction, $\mu_{\rm mol} \equiv M_{\rm mol}/M_*$;
(9) Gas depletion time ($M_{\rm mol}/$SFR).
\end{tabnote}
\end{table*}

\section{Molecular Gas Properties} \label{sec:gas}\label{sec:discussion}
In this section, we compare the molecular gas properties of FRB hosts with those of other galaxy populations and discuss the implications for FRB progenitor environments.

\subsection{Molecular Gas Mass}
We derived molecular gas masses from the CO observations. 
The CO line luminosity, $L'_{\rm CO}$, was calculated following \citet{solo05}: 
\begin{eqnarray}
L'_{\rm CO} = 3.25 \times 10^7 S_{\rm CO}\Delta v \nu_{\rm obs}^{-2} D_{\rm L}^2 (1+z)^{-3}, 
\end{eqnarray}
where $L'_{\rm CO}$ is in K~km~s$^{-1}$~pc$^2$, $S_{\rm CO}\Delta v$ is the velocity-integrated line flux in Jy~km~s$^{-1}$, $\nu_{\rm obs}$ is the observed CO line frequency in GHz, and $D_{\rm L}$ is the luminosity distance in Mpc. 
For the CO(2--1) observations, we estimated the CO(1--0) luminosity by assuming a CO line luminosity ratio of $r_{21} \equiv L'_{\rm CO(2-1)}/L'_{\rm CO(1-0)} = 0.65$, a typical value for nearby star-forming galaxies \citep{sain17,yaji21,den21}. 
For CO non-detections, 3$\sigma$ upper limits were derived by assuming a representative line width of 160 km~s$^{-1}$, corresponding to the average CO line width of the 
extended CO Legacy Data base for the GASS survey (xCOLD GASS; \cite{sain17})
sample over the stellar-mass range of the CO-undetected FRB hosts. 
The total molecular gas mass was derived using the CO-to-molecular-gas conversion factor, $\alpha_{\rm CO}$: 
\begin{eqnarray}
M_{\rm mol} = \alpha_{\rm CO} L'_{\rm CO(1-0)}. 
\end{eqnarray}
We adopted a Milky Way-like conversion factor of $\alpha_{\rm CO} = 4.3$ $M_{\odot}$~(K~km~s$^{-1}$~pc$^2$)$^{-1}$ \citep{bola13}, which includes the mass contribution from helium. 
This choice is appropriate for most of our targets because their gas-phase metallicities are in the range $12+\log({\rm O/H})=8.7$--$9.0$, except for the hosts of FRBs 20210807D and 20211127I with $12+\log({\rm O/H})=8.43$ and 8.44, respectively \citep{hein20a,gord23}. 
If the metallicity-dependent $\alpha_{\rm CO}$ prescription of \citet{genz15} were applied to these two hosts, their conversion factors would increase by a factor of $\sim$1.5, and their molecular gas masses would increase by the same factor. 
The derived molecular gas masses are summarized in Table~\ref{tab:results}. 
For the CO-detected and tentatively detected hosts, the molecular gas masses span $2.2\times10^8$--$1.0\times10^{10}$ $M_\odot$. 
We define the molecular gas fraction as $\mu_{\rm mol} \equiv M_{\rm mol}/M_*$, the ratio of molecular gas mass to stellar mass.

\subsection{Comparison with Stellar Mass and SFR}\label{sec:mgas-sfr}

\begin{figure}
\includegraphics[width=\linewidth]{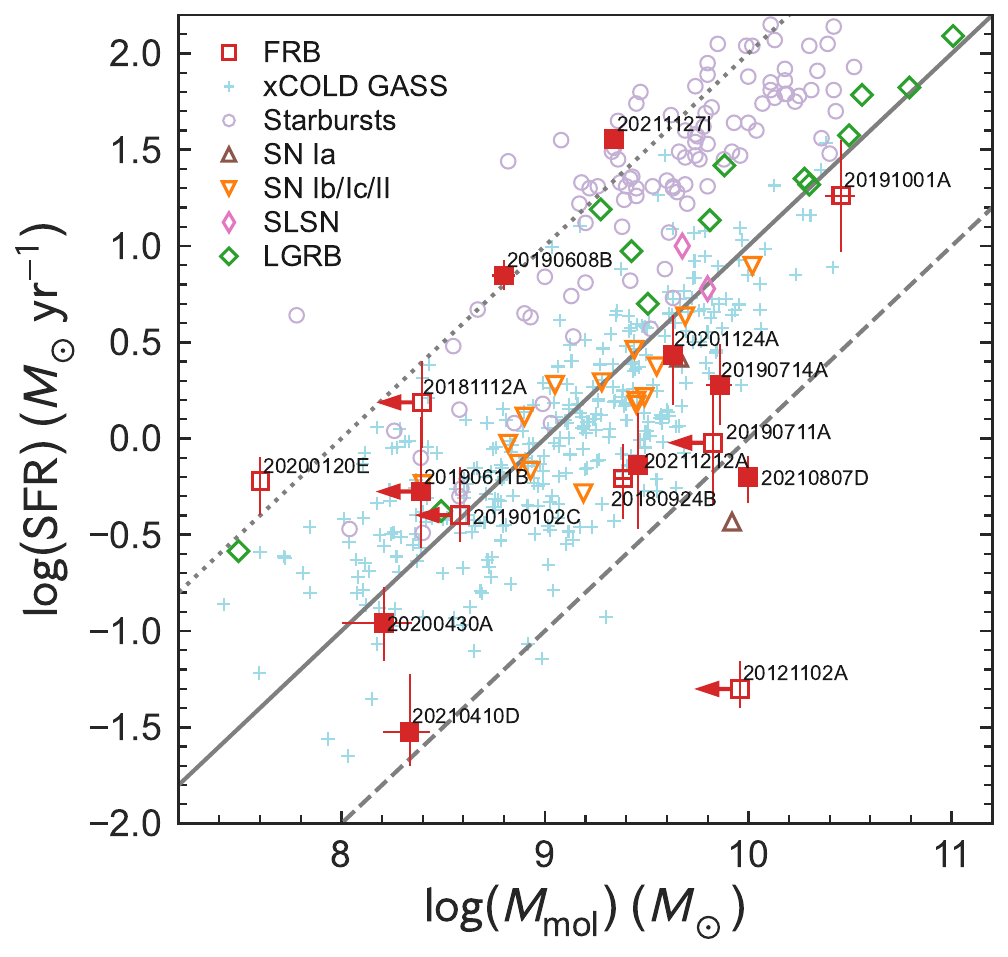}
\caption{
Comparison of molecular gas mass with SFR of the FRB hosts. 
The FRB hosts observed in this work are shown by filled red squares, and FRB hosts from the literature are shown by open red squares. 
Since the hosts of FRBs 20121102A and 20190711A have low metallicities ($12 + \log{\rm (O/H)} \le 8.4$; \cite{hein20a}), the conversion factor $\alpha_{\rm CO}$ is set to 200 and 60, respectively, following previous studies \citep{hats22,yama24a}, to set conservative upper limits on molecular gas mass. 
For comparison, we plot 
local galaxies from xCOLD GASS \citep{sain17},
local starburst galaxies \citep{kenn21}, 
SN hosts \citep{galb17}, 
SLSN hosts \citep{arab19,hats20a}, and 
long-duration GRB hosts \citep{hats20}. 
The dotted, solid, and dashed lines represent gas depletion times of 0.1, 1, and 10 Gyr, respectively.
}
\label{fig:mgas-sfr}
\end{figure}

We compare the molecular gas content of FRB host galaxies with their SFRs in Figure~\ref{fig:mgas-sfr}. 
The FRB sample includes the hosts observed in this study and those with molecular gas measurements or limits from previous studies \citep{bowe18,hats22,chit23,yama24a}. 
For the FRB\,20200430A host, the molecular gas mass is derived from the CO(3--2) line tentatively detected by \citet{chit23}, assuming a CO line luminosity ratio of $r_{31} \equiv L'_{\rm CO(3-2)}/L'_{\rm CO(1-0)} = 0.55$, a typical value for star-forming galaxies \citep{lamp20,cari13a}. 
Stellar masses and SFRs are taken from \citet{gord23}, except for the FRB\,20200120E host, for which we use the values from \citet{hein20a} because \citet{gord23} do not report these quantities for this host. 
Although FRB\,20200120E is shown in the figures for reference, it is excluded from the following discussion because it is the nearest known extragalactic FRB, is localized to a globular cluster associated with M81 \citep{bhar21,kirs22}, and may not belong to the same population as cosmological FRBs (e.g., \cite{nimm22,zhan24}).

As a local comparison sample for the statistical analyses below, we use xCOLD GASS \citep{sain17}. 
xCOLD GASS is a mass-selected CO(1--0) survey designed to provide a representative census of molecular gas in the local universe.
Its targets have $M_* > 10^9$ $M_\odot$ and $0.01 < z < 0.05$, and were randomly drawn from a parent sample of SDSS galaxies. 
The sample comprises 532 galaxies observed with the IRAM 30 m telescope. 
For comparison in Figure~\ref{fig:mgas-sfr}, we show xCOLD GASS together with local starburst galaxies \citep{kenn21} and host galaxies of other extragalactic transients, including supernovae (SNe) \citep{galb17}, super-luminous SNe (SLSNe) \citep{arab19,hats20a}, and long-duration gamma-ray bursts (LGRBs) \citep{hats14,hats20}.
Figure~\ref{fig:mgas-sfr} shows that the FRB hosts occupy a broad region of the $M_{\rm mol}$--SFR plane, rather than following a single locus. 
Several hosts overlap with the main distribution of xCOLD GASS galaxies, indicating that some FRB hosts have molecular gas reservoirs and star-formation activity consistent with those of ordinary local star-forming galaxies. 
At the same time, the FRB hosts extend toward both high and low star-formation efficiencies. 
The observed FRB hosts include both starburst-like systems and gas-rich systems with low star-formation efficiency. 
The molecular gas depletion time, defined as $\tau_{\rm depl}=M_{\rm mol}/{\rm SFR}$, ranges from 0.061 Gyr for the FRB\,20211127I host to 16 Gyr for the FRB\,20210807D host among the CO-detected hosts. 
The hosts of FRBs 20190608B and 20211127I lie close to the sequence of short depletion time comparable to those of nearby starburst galaxies. 
In contrast, the FRB\,20210807D host lies well below the main xCOLD GASS locus in SFR at its molecular gas mass, implying inefficient conversion of molecular gas into stars. 
This host has a large molecular gas reservoir but a long depletion time, suggesting that a gas-rich host does not necessarily have an elevated global SFR. 
Compared with the host galaxies of other transients shown in Figure~\ref{fig:mgas-sfr}, the FRB hosts are not confined to the region occupied by any single transient class. 
Several FRB hosts overlap with the SN, SLSN, and LGRB host samples, whereas some gas-rich FRB hosts have lower SFRs and longer depletion times than transient hosts with comparable molecular gas masses. 
Because these literature samples are heterogeneous in selection, redshift, CO transition, and sensitivity, they are used only as visual references and do not support a quantitative comparison between transient classes. 

This broad distribution was noted in previous studies \citep{hats22,yama24a}, and the addition of the present ALMA observations makes the diversity more apparent. 
The number of repeating FRB hosts in this comparison remains small (FRBs 20121102A, 20190711A, and 20201124A), so we do not attempt to compare repeating and apparently non-repeating FRB hosts separately.

\subsection{Statistical Assessment of Molecular Gas Properties}\label{sec:properties}

In this section, we test whether FRB host galaxies have molecular gas properties that differ from those of typical star-forming galaxies. 
The visual comparison in Figure~\ref{fig:mgas-sfr} shows a broad distribution of FRB hosts, but such a comparison alone does not account for the dependence of molecular gas content on stellar mass, SFR, and other host-galaxy properties. 
We therefore compare FRB hosts with controlled samples and scaling relations by using the following three complementary approaches. 

\subsubsection{Matched-control comparison for CO detections}\label{sec:matched-control}
We first construct matched control samples from xCOLD GASS to compare each FRB host with local galaxies having similar stellar mass and SFR. 
For each FRB host, control galaxies are selected within $|\Delta \log M_*| < 0.2$ and $|\Delta \log {\rm SFR}| < 0.3$. 
When fewer than 10 control galaxies satisfy these criteria, we instead adopt the 20 nearest neighbours in the normalized $\log M_*$--$\log {\rm SFR}$ plane. 
The nearest-neighbour distance is normalized by 0.2 dex in stellar mass and 0.3 dex in SFR, so that these differences are treated as comparable matching scales. 
These tolerances provide a practical balance between similarity in host properties and the number of available control galaxies. 
The molecular gas masses of xCOLD GASS galaxies are recalculated from their CO(1--0) luminosities using the same conversion factor as adopted for the FRB hosts, $\alpha_{\rm CO}=4.3$ $M_\odot$~(K~km~s$^{-1}$~pc$^2$)$^{-1}$. 
For the descriptive comparison below, the control medians are calculated from the CO-detected control galaxies. 
The effects of non-detections in both the FRB and control samples are assessed separately with survival analysis in the following subsection.
The hosts of FRBs\,20121102A and 20211127I lie well outside the stellar-mass--SFR parameter space covered by xCOLD GASS. 
We therefore exclude these two hosts from the matched-control statistics.

\begin{figure*}[ht]
\begin{center}
\includegraphics[width=.99\linewidth]{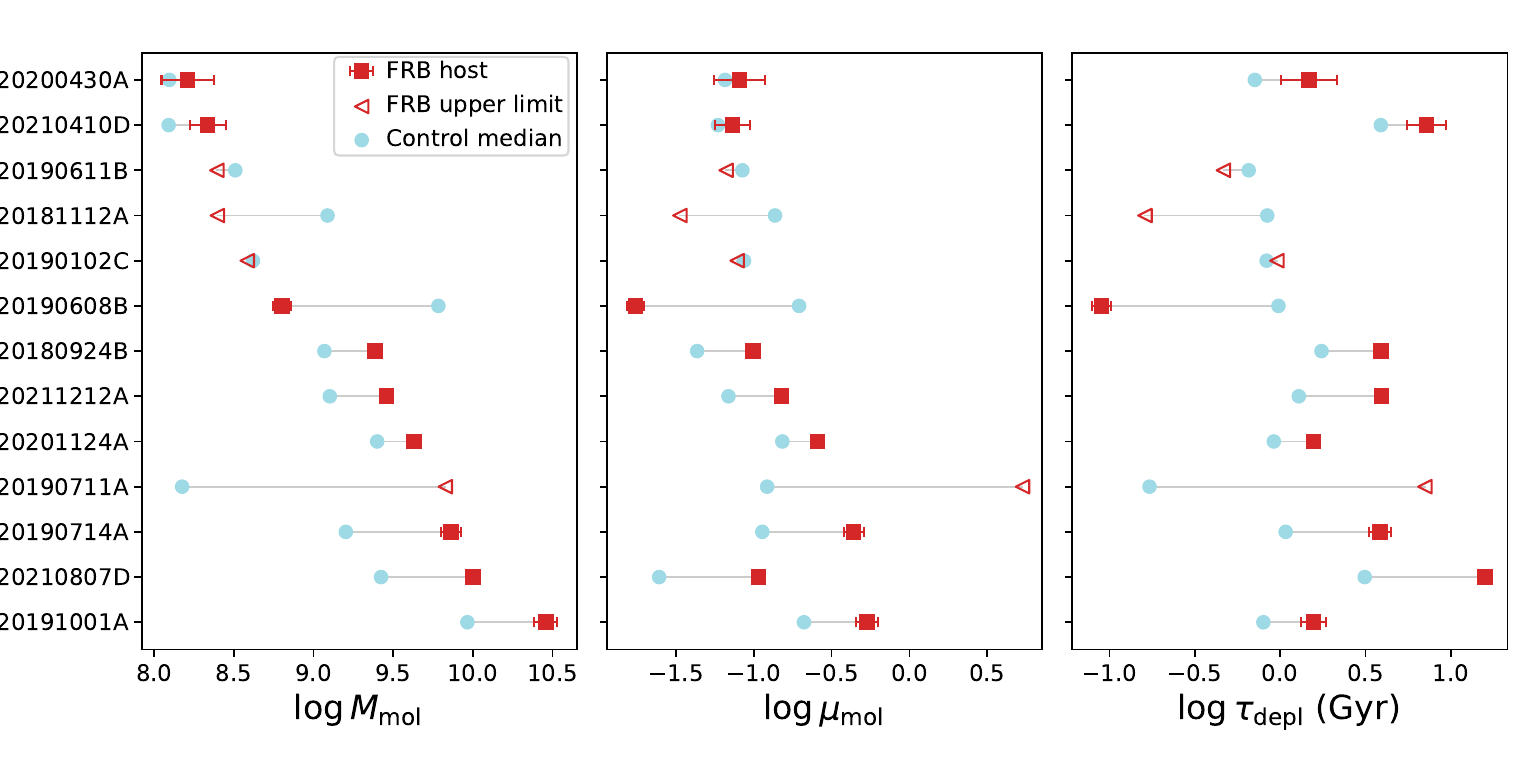}    
\end{center}
\caption{
Comparison between the 13 FRB hosts retained for the matched-control analysis and their matched control galaxies selected from xCOLD GASS. 
The three panels show molecular gas mass, molecular gas fraction, and molecular gas depletion time. 
Squares show CO-detected FRB hosts, open triangles show upper limits for CO non-detections, and blue circles show the detection-only median values of the matched-control samples. 
Horizontal error bars show the measurement uncertainties for the FRB hosts. 
Horizontal gray lines connect each FRB host to its matched-control median.
}
\label{fig:matched-controls}
\end{figure*}

Figure~\ref{fig:matched-controls} compares the remaining 13 FRB hosts with the median values of their matched controls. 
Based on their central values, eight of the nine FRB hosts represented by CO detections in this comparison, including the tentative detections, have larger molecular gas masses than their matched-control medians. 
The median offset is $\Delta \log M_{\rm mol}=+0.32$ dex, corresponding to a factor of $\sim$2.1 enhancement. 
The molecular gas fractions show a similar trend: 8 out of 9 CO-detected FRB hosts have $\mu_{\rm mol}$ above the matched-control median, with a median offset of $\Delta \log \mu_{\rm mol}=+0.34$ dex, or a factor of $\sim$2.2. 
The depletion times are also generally longer than those of the controls, with 8 out of 9 CO-detected FRB hosts showing positive offsets and a median $\Delta \log \tau_{\rm depl}=+0.32$ dex. 
We verified that by adopting equal normalization scales of 0.2 dex or 0.3 dex for both stellar mass and SFR, the conclusion that FRB hosts tend to be molecular-gas rich relative to matched local galaxies is robust to this choice. 

Within this detection-only comparison, the offsets suggest that many FRB hosts contain larger molecular gas reservoirs and longer gas depletion time than local galaxies with similar stellar masses and SFRs. 
At the same time, the FRB host population remains diverse: the measurements or upper limits for the hosts of FRBs 20190611B, 20181112A, 20190102C, and 20190608B are comparable to or lower than the corresponding control medians.

\subsubsection{Survival analysis including CO non-detections}
We next directly compare the molecular gas properties of the 13 FRB hosts retained in the matched-control analysis with those of the corresponding xCOLD GASS galaxies. 
FRBs\,20121102A and 20211127I are excluded because no xCOLD GASS galaxies satisfy the matching criteria, as described in Section~\ref{sec:matched-control}. 
The matched-control sets contains 222 unique xCOLD GASS galaxies. 
The FRB and control samples contain nine detections and four upper limits, and 184 detections and 38 upper limits, respectively. 
The CO non-detections in both samples are treated as left-censored measurements. 
To apply standard right-censoring methods, we reverse the signs of the measured values and upper limits, following the treatment of upper limits in survival analysis \citep{feig85}. 
The reverse Kaplan--Meier estimator provides non-parametric estimates of the distributions and their medians \citep{kapl58,feig85}. 

Survival analysis of molecular gas in FRB hosts was previously performed by \citet{hats22}, who applied the Kaplan--Meier estimator and log-rank test to six FRB hosts and the local galaxy sample. 
They found a possible difference in molecular gas fraction, but no difference in depletion time. 
The molecular gas fraction result was sensitive to the extremely gas-poor environment of FRB\,20200120E in an M81 globular cluster: after excluding this object, the difference became inconclusive. 
The present analysis extends that work by increasing the test sample to 13 FRB hosts and by comparing them with xCOLD GASS controls selected individually to have similar stellar masses and SFRs. 
It also applies a common CO-to-molecular-gas conversion to the FRB and control samples and excludes the Galactic and M81 events, thereby reducing sensitivity to a single extreme nearby environment and to differences in the host-property distributions. 

We use two tests to compare the distributions while retaining the information provided by CO upper limits, rather than discarding non-detections or treating their limits as measured values.
We adopt the Peto--Peto test as the primary test. It weights the contributions at each detected value using the pooled Kaplan--Meier survival probability \citep{peto72}.
The Peto--Peto test places less emphasis on the low values that are difficult to constrain with the available upper limits, although those limits still leave uncertainty in the distribution.
For comparison, we also use the log-rank test, which does not apply this additional weighting toward high values.
Comparing the two results helps us check whether an apparent difference between FRB hosts and control galaxies depends on which part of the distribution receives more weight.
Both tests assume that, within each sample, the detection limits are independent of the true gas-property values as this is the assumption of non-informative censoring.

The Peto--Peto and log-rank tests for $M_{\rm mol}$ give $p=0.450$ and $p=0.813$, respectively, providing no evidence that the two distributions differ. 
The Peto--Peto probabilities for $\mu_{\rm mol}$ and $\tau_{\rm depl}$ are $p=0.345$ and $0.058$, respectively, while the corresponding log-rank probabilities are $p=0.491$ and $0.668$. 
Thus, neither $M_{\rm mol}$ nor $\mu_{\rm mol}$ differs significantly between the FRB hosts and matched controls when CO non-detections are included. 
The Peto--Peto result for $\tau_{\rm depl}$ suggests a marginal tendency toward longer depletion times in FRB hosts. 

\subsubsection{Comparison with molecular-gas scaling relations}
Finally, since the molecular gas mass fraction and depletion time depend on redshift, stellar mass, and offset from the main sequence (MS) of star-forming galaxies \citep{genz15,tacc18}, we compare the molecular gas properties with scaling relations that account for these parameters. 
\citet{tacc18,tacc20} parameterize the molecular gas fraction and depletion time as functions of redshift, stellar mass, and the offset from the star-forming main sequence. 
The main-sequence offset is defined as $\Delta_{\rm MS}=\log({\rm SFR}/{\rm SFR}_{\rm MS})$, where ${\rm SFR}_{\rm MS}$ is evaluated at the redshift and stellar mass of each FRB host using the main-sequence prescription of \citet{spea14}.

\begin{figure*}[ht]
\includegraphics[width=0.49\linewidth]{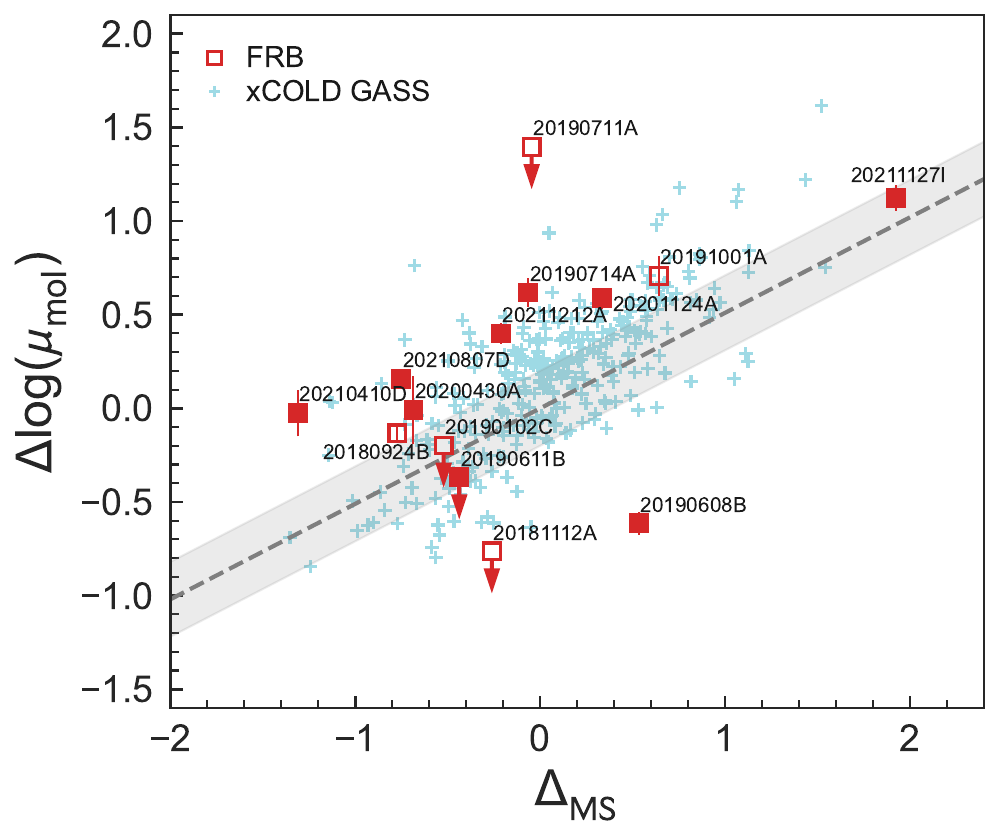}
\hspace{2mm}
\includegraphics[width=0.49\linewidth]{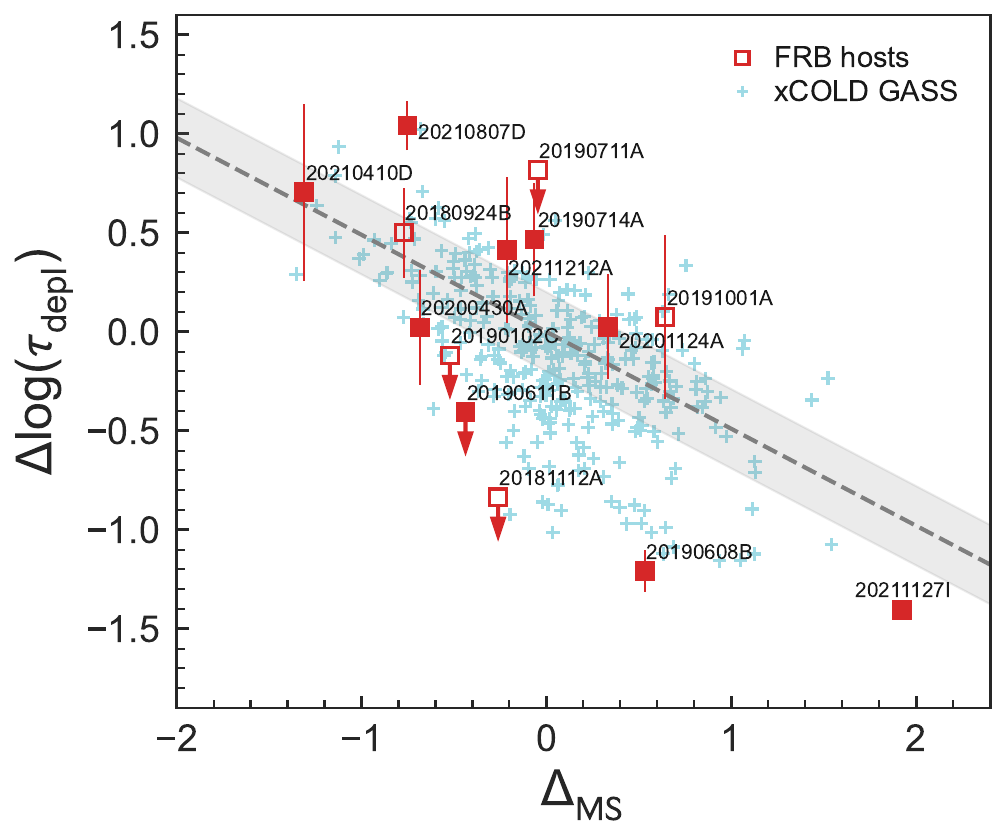}
\caption{
Redshift-corrected molecular gas fraction (left) and depletion time (right) as functions of the offset from the star-forming main sequence.
Red squares show FRB hosts. 
Filled symbols indicate hosts observed in this work, and downward arrows indicate upper limits. 
Blue crosses show xCOLD GASS \citep{sain17}. 
The dashed curves show the \citet{tacc20} scaling relations after the same redshift correction, with shaded regions showing a representative $\pm0.25$ dex scatter for individual galaxies.
}
\label{fig:delta-ms-gas}
\end{figure*}

Figure~\ref{fig:delta-ms-gas} shows the redshift-corrected logarithmic molecular gas fraction and depletion time as functions of $\Delta_{\rm MS}$ for the FRB hosts and xCOLD GASS galaxies.
In both panels, the redshift dependence of the \citet{tacc20} scaling relations is removed.
The dashed curves show the \citet{tacc20} scaling relations after the same redshift correction, and the shaded regions indicate a representative scatter of $\pm0.25$ dex for individual galaxies, comparable to the $0.2$--$0.25$ dex scatter \citep{tacc18}.
Seven of the ten hosts represented by CO detections have positive residuals in both quantities.
The CO-detected FRB hosts show median residuals of $+0.40$ dex and $+0.44$ dex for gas fraction and depletion time, respectively, indicating a tendency for CO-detected FRB hosts to have enhanced molecular gas fractions and longer depletion times.
The upper limits should not be interpreted as secure positive offsets, and the present results support diversity in FRB host molecular-gas properties rather than a uniformly gas-rich population.

\subsection{Atomic and molecular gas}\label{sec:HI}

Atomic and molecular gas trace complementary components of the cold interstellar medium. 
Their combined mass measures the available cold-gas reservoir, while their mass ratio describes the partition between the atomic and molecular phases. 
Two hosts in our sample, FRBs\,20211127I and 20211212A, have detections of both CO and \textsc{Hi} \citep{glow23, roxb25}. 
Their \textsc{Hi} emission extends over 40 kpc, substantially beyond the molecular-gas and stellar distributions. 
The corresponding \textsc{Hi} masses are $\log(M_\textsc{Hi}/M_\odot)=9.99$ and 9.65, respectively \citep{roxb25}. 

\begin{figure*}[ht]
\includegraphics[width=0.49\linewidth]{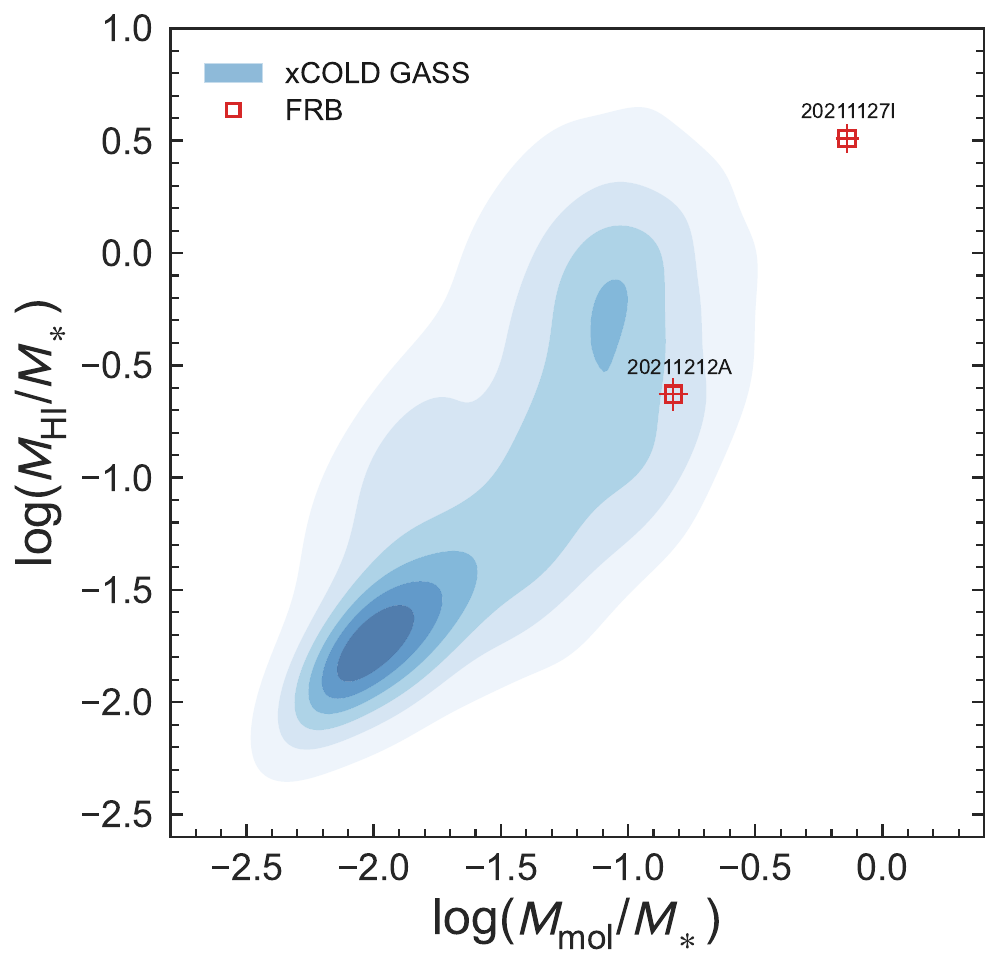}
\hspace{2mm}
\includegraphics[width=0.49\linewidth]{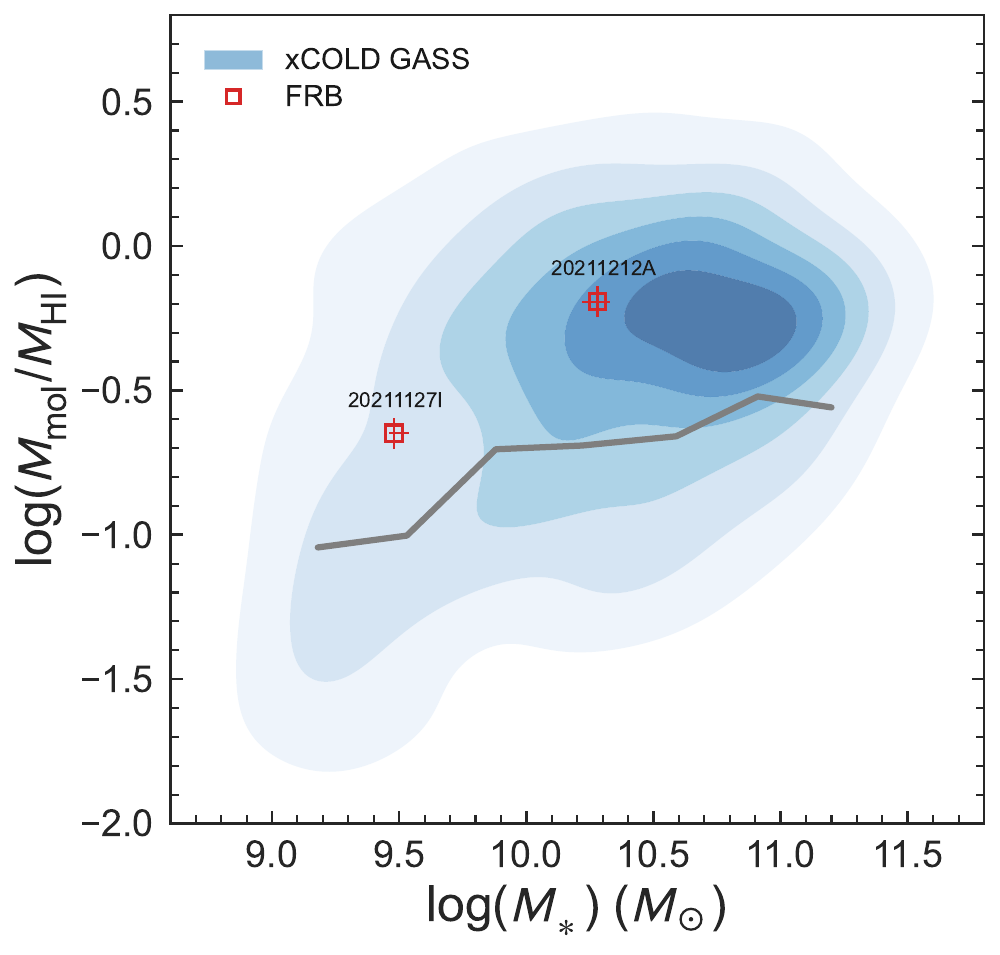}
\caption{
Left: molecular gas fraction versus atomic gas fraction for FRB hosts detected in both CO and \textsc{Hi}. 
Right: molecular-to-atomic gas mass ratio as a function of stellar mass. 
Red open squares show the hosts of FRBs\,20211127I and 20211212A. 
Blue shaded density contours show the matched-control subset of galaxies jointly observed by xGASS and xCOLD GASS \citep{sain17,cati18} and detected in both CO and \textsc{Hi}.
The gray curve in the right panel shows the running median of the full xCOLD GASS sample.
}
\label{fig:atomic-molecular}
\end{figure*}

Figure~\ref{fig:atomic-molecular} compares their atomic and molecular gas fractions and their molecular-to-atomic gas mass ratios with the joint sample from
the extended \textit{GALEX} Arecibo SDSS Survey (xGASS; \cite{cati18})
and xCOLD GASS. 
This comparison sample is drawn from the stellar-mass-selected xGASS survey of nearby galaxies with integrated \textsc{Hi} measurements and its xCOLD GASS subsample observed in CO(1--0), providing a reference set with measurements of both the atomic and molecular gas phases over $10^9 < M_*/M_\odot < 10^{11.5}$ \citep{sain17,cati18}. 
The blue shaded density contours in Figure~\ref{fig:atomic-molecular} show the matched-control subset detected in both tracers.
In contrast, the gray curve in the right panel shows the running median of the full xCOLD GASS sample.
The FRB\,20211127I host is a gas-rich outlier in both phases, placing it well outside the main distribution of the comparison sample in the gas-fraction plane. 
On the other hand, its molecular-to-atomic ratio ($R_{\rm mol}=M_{\rm mol}/M_\textsc{Hi}$) is $\log R_{\rm mol}=-0.65$, and its large molecular gas mass primarily reflects an exceptionally large total cold-gas reservoir rather than a molecular-dominated gas budget. 
The FRB\,20211212A host has less extreme gas fractions and lies within the broad locus of the comparison sample. 
Its molecular-to-atomic ratio is higher than that of the FRB\,20211127I host ($\log R_{\rm mol}=-0.19$). 
At their respective stellar masses, both FRB hosts lie above the running median of the full xCOLD GASS sample, although their ratios remain within the broad scatter of the matched-control sample.

This comparison suggests that a large molecular gas content in an FRB host does not necessarily imply an unusually efficient conversion of \textsc{Hi} into molecular gas. 
Because only two FRB hosts currently have detections in both phases, and because requiring both CO and \textsc{Hi} detections introduces a gas-rich selection bias, these results should be regarded as individual case studies rather than evidence for a common property of the FRB host population. 
A larger, uniformly selected sample with both CO and \textsc{Hi} measurements is required to distinguish systematically between enhanced total gas reservoirs and altered atomic-to-molecular gas partitioning.

\subsection{Implications for the formation environment of FRB progenitors}\label{sec:implications}

The molecular gas properties measured in this work provide constraints on the galactic environments in which FRB progenitors form. 
In the detection-only comparisons, molecular gas fractions and depletion times tend to be larger than those of matched galaxies and the values expected from scaling relations for normal star-forming galaxies. 
However, the survival analysis including CO non-detections does not establish a statistically significant population-wide enhancement. 
Molecular-gas richness should therefore be regarded as a property of a subset of FRB hosts rather than a necessary condition for producing an FRB source. 
The long depletion times of several hosts further indicate that a large gas reservoir is not accompanied by a proportionally high current SFR. 
Such systems may have recently accumulated gas, may convert gas into stars inefficiently, or may contain molecular gas that is dynamically stabilized against collapse.

The prevalence of star-forming FRB hosts is broadly consistent with a prompt formation channel in which magnetars are produced by the core collapse of massive stars. 
Stellar-population analyses show that most localized FRBs occur in star-forming galaxies, while a smaller number arise in transitioning or quiescent systems, consistent with a substantial short-delay contribution together with delayed channels \citep{gord23}. 
The deficit of low-mass FRB hosts reported by \citet{shar24} may imply that FRB production is biased toward massive, metal-rich star-forming galaxies, possibly because metal-rich binary evolution facilitates stellar mergers that can produce strongly magnetized stellar cores. 
On the other hand, direct comparisons of host metallicities have not found clear evidence for a metallicity preference \citep{yama26}. 
Our finding that some FRB hosts are rich in molecular gas is compatible with the formation of young magnetars, because such reservoirs can sustain recent or renewed star formation, but it does not by itself demonstrate a metallicity-dependent or massive-star origin.

The CO velocity fields provide an additional constraint on whether the gas-rich environments are associated with dynamically disturbed systems. 
All six securely CO-detected hosts with sufficient spatially resolved emission show large-scale velocity gradients (Figure~\ref{fig:mom1}). 
The hosts of FRBs 20201124A, 20211127I, and 20211212A exhibit relatively smooth gradients consistent with ordered disk rotation, while the compact CO component in the FRB\,20190608B host also shows a systematic gradient. 
The velocity fields of the FRB\,20190714A and FRB\,20210807D hosts are more patchy and asymmetric, but their clumpy CO distributions and locally low surface brightness prevent the irregularities from being uniquely attributed to non-circular motions or interactions. 
The diversity in molecular gas depletion time therefore does not correspond to a simple distinction between ordered and disturbed kinematics. 
Moreover, none of the hosts shows a clear velocity discontinuity or separate kinematic component specifically at the FRB position, although the present kpc-scale resolution cannot identify smaller-scale motions. 

Gas accretion and galaxy interactions nevertheless provide one possible mechanism for producing the large cold-gas reservoirs observed in some hosts. 
Previous \textsc{Hi} studies identified asymmetric profiles, tidal structures, or nearby gas-rich companions in individual FRB hosts \citep{mich21,kaur22}. 
In molecular gas, an asymmetric velocity gradient has been reported in the FRB\,20180924B host, whereas the FRB\,20191001A host shows a rotating CO disk \citep{hsu23,yama24a}. 
These observations motivated scenarios in which an interaction supplies or compresses gas, triggers localized star formation, and forms a short-lived FRB progenitor. 
In the present sample, however, the predominance of ordered large-scale velocity gradients and the absence of unambiguous major kinematic disturbances do not support recent strong interactions as a common property. 
The atomic--molecular comparison in Section~\ref{sec:HI} shows that gas-rich FRB hosts do not need to share the same gas balance: FRB\,20211127I is dominated by an exceptionally large atomic reservoir, whereas FRB\,20211212A has a larger fraction of its cold gas in the molecular phase. 
Moreover, the expanded \textsc{Hi} host sample of \citet{roxb25} does not show a statistically significant excess of disturbed morphologies relative to normal galaxies. 
Interactions or recent gas accretion may therefore be relevant for individual hosts, but neither the molecular- nor atomic-gas observations support them as a universal requirement for FRB progenitor formation.

An additional limitation is that integrated gas measurements characterize the host galaxy rather than the immediate FRB environment. 
In the host sample studied by \citet{mann21c}, FRBs broadly follow the radial distribution of host-galaxy light but are not preferentially located in regions with the highest local star-formation or stellar-mass surface densities. 
Using a larger sample and comparing FRB locations with host-galaxy light, spiral-arm substructure, and modeled globular-cluster distributions, \citet{gord25} found that most FRBs favor locations in galactic disks, while only $11\%\pm5\%$ favor a globular-cluster origin. At $z<0.15$, approximately 20\%--46\% favor an association with spiral arms, consistent with a diverse population in which multiple progenitor channels may contribute.
In addition, the localization of FRB\,20200120E to an old globular cluster demonstrates that at least some FRB sources can form through delayed or dynamical channels without a young, gas-rich environment \citep{kirs22}. 
The coexistence of gas-rich star-forming hosts, less active hosts, and an old globular-cluster environment favors either multiple progenitor channels or a single source class, such as magnetars, that can be produced through several evolutionary pathways.

Only a limited number of hosts show significant CO emission at the FRB position at the sensitivity and spatial resolution of the available observations. 
For the remaining hosts, including the local upper limits does not reveal a clear common trend in molecular gas content at the burst sites. 
The most direct test of a connection between cold gas and FRB progenitors will therefore require higher-resolution spatially resolved observations. 
Sub-kpc CO and \textsc{Hi} imaging should determine whether the apparently ordered velocity fields remain regular on smaller scales and whether FRB positions coincide with molecular clouds, high gas surface densities, recent gas inflows, or local kinematic disturbances. 
Combining these maps with resolved stellar ages, metallicities, extinction, and recent star-formation histories would distinguish a progenitor formed in situ from one formed earlier and subsequently displaced. 
Larger and uniformly selected CO and \textsc{Hi} samples, including non-detections, are also required to establish whether gas-richness, gas-phase partitioning, or disturbed kinematics occur more frequently in FRB hosts than in galaxies matched in stellar mass, SFR, and redshift.


\section{Radio emission in the FRB\,20201124A host}\label{sec:prs}

A persistent radio source (PRS) has been reported in association with FRB\,20201124A, a repeating FRB at $z = 0.0979$ \citep{piro21a,fong21,ravi22b}. 
Previous studies have generally interpreted the persistent radio emission as being dominated by star formation in the host galaxy. 
VLA 22 GHz imaging resolved the emission on an arcsecond scale at a position consistent with the FRB within $0\farcs2$ \citep{piro21a}.
Based on this result and the non-detection of a compact source in 
the European VLBI Network (EVN; \cite{schi95}) and the enhanced Multi-Element Remotely Linked Interfer- ometer Network (\textit{e}-MERLIN; \cite{garr04})
observations on 2--100 mas scales, \citet{piro21a} argued that the radio emission is powered by enhanced star formation near the FRB. 
Optical and infrared studies have also shown that FRB\,20201124A resides in a dusty, star-forming spiral host, with the burst position lying in or near a star-forming region \citep{fong21,dong24a}. 
Similarly, \citet{ravi22b} identified a persistent radio counterpart and found that the emission is extended on scales of $\gtrsim50$ mas, favoring an origin related to ongoing star formation. 
Our CO detection provides an independent tracer of the cold gas reservoir that can fuel such star formation. 
The overlap between the CO-emitting host, the ALMA continuum emission, and the previously reported radio emission therefore supports a picture in which at least part of the persistent radio emission is linked to star-forming activity in a molecular-gas-rich region.

A different view was presented by \citet{brun24}, who interpreted the persistent radio emission associated with FRB\,20201124A in the context of a synchrotron nebula powered by a young magnetar. 
They argued that a star-formation origin is disfavored because the radio source is localized to the FRB position rather than being distributed over the host galaxy, and because the local SFR inferred from high-resolution optical H$\alpha$ and millimeter dust-continuum measurements is too low to account for the observed persistent radio luminosity. 
They further showed that FRB\,20201124A extends the empirical relation between PRS luminosity and Faraday rotation measure to lower luminosities and lower rotation measures than the brighter PRSs associated with FRB\,20121102A and FRB\,20190520B, supporting a compact magneto-ionic nebula physically connected to the FRB source. 
They also noted that the radio spectral index measured from VLA observations at 6, 15, and 22 GHz is opposite in sign to those measured for the PRSs associated with FRB\,20121102A and FRB\,20190520B. 
Extrapolating this spectrum to 100 GHz yields an expected flux density of $\sim$40--500 $\mu$Jy. 

\begin{figure}
\includegraphics[width=\linewidth]{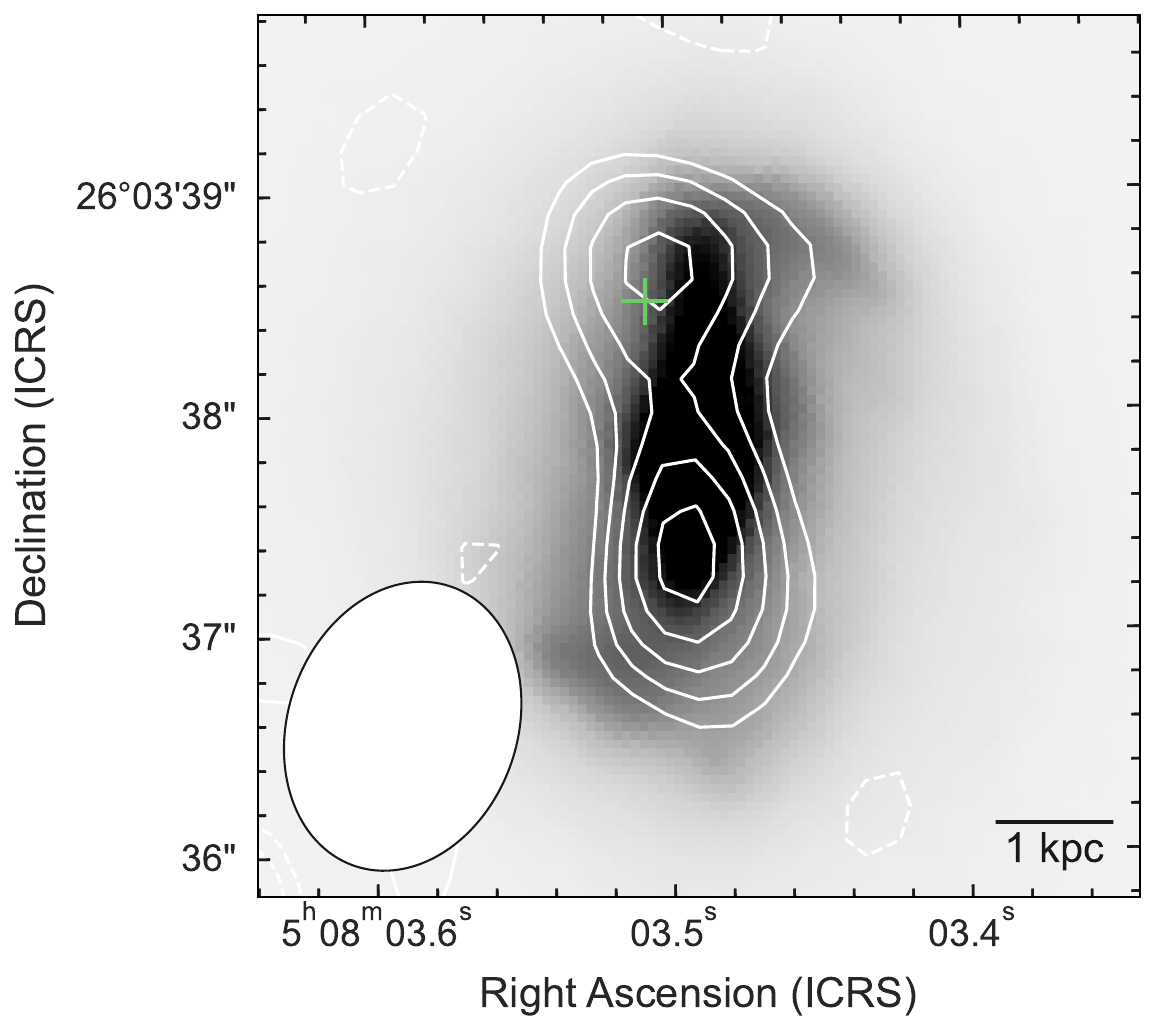}
\caption{
Contours of ALMA 98 GHz (3.1 mm) continuum overlaid on optical image of the FRB\,20201124A host. 
The continuum map is created with a robust parameter of 0.5. 
The synthesized beam size ($1\farcs3 \times 1\farcs0$) is shown in the lower-left corner. 
The contours start at $\pm1.5\sigma$ and increase in steps of $0.5\sigma$; negative contours are shown by dashed lines. 
The FRB\,20201124A is localized with milliarcsecond precision and its position is indicated with a cross mark.
}
\label{fig:prs}
\end{figure}

Figure~\ref{fig:prs} shows the contours of the ALMA 98 GHz (3.1 mm) continuum map overlaid on the optical image of the FRB\,20201124A host galaxy. 
The continuum map here was created with a robust parameter of 0.5, yielding a synthesized beam size of $1\farcs3 \times 1\farcs0$. 
The continuum emission shows two peaks: the northern peak is coincident with the FRB position, while the southern peak is closer to the galaxy center. 
This spatial distribution is consistent with the VLA 15 GHz observations presented by \citet{brun24}.

We examined whether the millimeter continuum emission around the FRB position (peak flux density $S_{\rm 3.1mm}=39\pm13$ $\mu$Jy) can be explained by star formation in the host galaxy. 
Following \citet{murp12a}, we estimated the radio continuum expected from star formation as the sum of non-thermal synchrotron and thermal free--free emission, adopting an electron temperature of $10^4$ K and a non-thermal spectral index of 0.8. 
At the observed frequency of 98 GHz, a local SFR of 0.4 $M_{\odot}$~yr$^{-1}$, estimated from the H$\alpha$ emission line \citep{brun24}, gives expected synchrotron and free--free emission of 0.5 and 1.9 $\mu$Jy, respectively. 
We also estimated the thermal dust contribution using a modified blackbody with dust temperatures of $T_{\rm dust}=20$ and 30 K, representative of nearby star-forming galaxies, and a dust emissivity index of $\beta=1.5$ \citep{gala12, cort14, lamp19}. 
The predicted dust emission is 19 $\mu$Jy for $T_{\rm dust}=20$ K and 3.1 $\mu$Jy for $T_{\rm dust}=30$ K. 
The total star-formation-related flux density including the synchrotron and free--free components is below the observed flux density. 
The ALMA flux density lies at the lower end of or below the 100 GHz flux density expected from a simple extrapolation of the VLA spectrum.
Using the ALMA and VLA data, \citet{brun26a} fitted a spectral peak at $\nu_{\rm m}=63^{+100}_{-33}$ GHz, consistent with a young magnetar wind nebula (MWN).
In the MWN interpretation of \citet{raha25}, PRS emission is synchrotron radiation from relativistic particles energized by a magnetar wind or by magnetic-field decay, and FRB\,20201124A is treated as a candidate PRS in this framework. 
The relatively faint ALMA 98 GHz flux is comparable to the flux level expected near the cooling break in the MWN model, and may indicate that the radio spectrum turns over between the VLA and ALMA bands rather than following a single power law to millimeter wavelengths.
The excess emission at the FRB position may include a compact component associated with the PRS or an MWN, although the present kpc-scale ALMA beam cannot isolate such a compact source from surrounding star-forming emission.
Deeper and higher-angular resolution millimeter and radio observations are required to test for such a turnover and to separate these components robustly.

\section{Conclusions}\label{sec:conclusions}

We presented ALMA CO(1--0) and CO(2--1) observations of nine localized FRB host galaxies at $z=0.05$--0.38, together with millimeter continuum measurements.
Combining these observations with molecular-gas measurements from previous studies, we compared the host molecular gas masses, gas fractions, and depletion times with stellar-mass- and SFR-matched xCOLD GASS galaxies and with molecular-gas scaling relations.
The combined sample of FRB hosts with CO measurements or upper limits is the largest among those analyzed in previous molecular-gas studies of FRB hosts.
We also examined the resolved CO distributions and kinematics, the atomic and molecular gas reservoirs where both tracers are available, and the origin of the millimeter continuum emission near FRB\,20201124A.
Our main findings are as follows.

\begin{itemize}
\item
CO emission is significantly detected in six of the nine hosts and tentatively detected in the FRB\,20210410D host.
The hosts of FRBs\,20190611B and 20200430A are undetected in our CO observations.
With the adopted Milky Way CO-to-H$_2$ conversion factor, the detected and tentatively detected hosts have molecular gas masses of $2.2\times10^8$--$1.0\times10^{10}$ $M_\odot$.
Millimeter continuum emission is tentatively detected in three hosts.

\item
The six significantly CO-detected hosts show spatially resolved emission and large-scale velocity gradients.
Several velocity fields are consistent with ordered disk rotation, and no clear velocity discontinuity or distinct kinematic component is identified at the FRB positions.
The FRB positions do not necessarily coincide with the brightest CO emission.
These results do not require strongly disturbed molecular-gas environments, although the present kpc-scale resolution cannot exclude local disturbances.

\item
In the detection-only matched-control comparison, eight of nine FRB hosts have molecular gas masses above their respective control medians.
The gas fractions and depletion times show similar trends.
The comparison with molecular-gas scaling relations likewise gives positive median offsets in gas fraction and depletion time for the CO-detected hosts.
Nevertheless, individual hosts span diverse gas properties, including measurements or upper limits comparable to or below the control medians.

\item
Including CO non-detections in both the FRB and matched-control samples changes the population-level interpretation.
Neither the Peto--Peto nor the log-rank test establishes a significant difference in molecular gas mass or gas fraction.
The marginal tendency toward longer depletion times is found in the Peto--Peto test ($p=0.058$).
Thus, the present data do not establish a population-wide molecular-gas enhancement or uniformly long depletion time among FRB hosts.

\item
The two hosts with both CO and \textsc{Hi} detections illustrate different cold-gas reservoirs.
The FRB\,20211127I host is exceptionally gas-rich in both phases but remains atomic-gas dominated, whereas the FRB\,20211212A host has less extreme gas fractions.
Both molecular-to-atomic mass ratios remain within the broad scatter of the comparison galaxies.
These individual cases do not establish unusual gas-phase partitioning as a general property of FRB hosts.

\item
The ALMA 98 GHz (3.1 mm) continuum map of the FRB\,20201124A host shows a peak coincident with the FRB position.
The predicted star-formation-related emission is below the measured flux density.
The ALMA flux density lies at the lower end of or below a simple extrapolation of the VLA spectrum.
A compact PRS or MWN contribution and a spectral turnover remain possible, but the current sensitivity and spatial resolution do not allow a firm separation from surrounding star-forming emission.
\end{itemize}

Larger, uniformly selected CO samples with well-characterized upper limits are needed to determine whether FRB hosts differ systematically from galaxies matched in stellar mass, SFR, and redshift.
Sub-kpc gas imaging combined with resolved stellar populations and recent star-formation histories will connect the host-scale reservoirs to the local FRB environments and help distinguish progenitor formation pathways.
For FRB\,20201124A, deeper radio and millimeter observations at higher angular resolution are required to isolate any compact persistent source and test for a spectral turnover.

\begin{ack}
We would like to acknowledge the reviewer for valuable comments and suggestions, which significantly improved the article. 
We would like to express our gratitude to the ALMA staff for data acquisition and support. 
BH is supported by JSPS KAKENHI Grant Numbers 19K03925 and 23K03449. 
TH acknowledges support from Taiwan's National Science and Technology Council (NSTC) through grants 113-2112-M-005-009-MY3, 114-2123-M-001-002-, and 111-2112-M-005-018-MY3.
T-YH acknowledges support from the Australian Research Council Centre of Excellence for Gravitational Wave Discovery (CE230100016).
KK acknowledges support from JSPS KAKENHI Grant Numbers JP22H04939, JP23K20035, and JP24H00004.
This work is supported by the ALMA Japan Research Grant of NAOJ Chile Observatory (NAOJ-ALMA-416).
This paper makes use of the following ALMA data: ADS/JAO.ALMA\#2021.1.00027.S and \#2023.1.01090.S. 
ALMA is a partnership of ESO (representing its member states), NSF (USA) and NINS (Japan), together with NRC (Canada), NSTC and ASIAA (Taiwan), and KASI (Republic of Korea), in cooperation with the Republic of Chile. The Joint ALMA Observatory is operated by ESO, AUI/NRAO and NAOJ.
This project used public archival data based on observations made with the NASA/ESA Hubble Space Telescope, and obtained from the Hubble Legacy Archive, which is a collaboration between the Space Telescope Science Institute (STScI/NASA), the Space Telescope European Coordinating Facility (ST-ECF/ESAC/ESA) and the Canadian Astronomy Data Centre (CADC/NRC/CSA).
This project used public archival data from the Dark Energy Survey (DES). Funding for the DES Projects has been provided by the U.S. Department of Energy, the U.S. National Science Foundation, the Ministry of Science and Education of Spain, the Science and Technology Facilities Council of the United Kingdom, the Higher Education Funding Council for England, the National Center for Supercomputing Applications at the University of Illinois at Urbana-Champaign, the Kavli Institute of Cosmological Physics at the University of Chicago, the Center for Cosmology and Astro-Particle Physics at the Ohio State University, the Mitchell Institute for Fundamental Physics and Astronomy at Texas A\&M University, Financiadora de Estudos e Projetos, Funda{\c c}{\~a}o Carlos Chagas Filho de Amparo {\`a} Pesquisa do Estado do Rio de Janeiro, Conselho Nacional de Desenvolvimento Cient{\'i}fico e Tecnol{\'o}gico and the Minist{\'e}rio da Ci{\^e}ncia, Tecnologia e Inova{\c c}{\~a}o, the Deutsche Forschungsgemeinschaft, and the Collaborating Institutions in the Dark Energy Survey.
The Collaborating Institutions are Argonne National Laboratory, the University of California at Santa Cruz, the University of Cambridge, Centro de Investigaciones Energ{\'e}ticas, Medioambientales y Tecnol{\'o}gicas-Madrid, the University of Chicago, University College London, the DES-Brazil Consortium, the University of Edinburgh, the Eidgen{\"o}ssische Technische Hochschule (ETH) Z{\"u}rich,  Fermi National Accelerator Laboratory, the University of Illinois at Urbana-Champaign, the Institut de Ci{\`e}ncies de l'Espai (IEEC/CSIC), the Institut de F{\'i}sica d'Altes Energies, Lawrence Berkeley National Laboratory, the Ludwig-Maximilians Universit{\"a}t M{\"u}nchen and the associated Excellence Cluster Universe, the University of Michigan, the National Optical Astronomy Observatory, the University of Nottingham, The Ohio State University, the OzDES Membership Consortium, the University of Pennsylvania, the University of Portsmouth, SLAC National Accelerator Laboratory, Stanford University, the University of Sussex, and Texas A\&M University.
Based in part on observations at Cerro Tololo Inter-American Observatory, National Optical Astronomy Observatory, which is operated by the Association of Universities for Research in Astronomy (AURA) under a cooperative agreement with the National Science Foundation.
\end{ack}

\bibliographystyle{aasjournal}
\bibliography{zotero}


\end{document}